\documentclass[aps,prd,a4paper,eqsecnum,%
superscriptaddress,nofootinbib,showpacs,%
twocolumn,showkeys,amsfonts,amssymb,amsmath]{revtex4}

\usepackage{ amsthm}

\newtheorem{theo}{Proposition}
\newtheorem{lemma}{Lemma}

\newcommand{\sgn}{\mathop {\rm sgn}\nolimits}
\newcommand{\sgt}{\frac{1}{\gamma}}
\newcommand{\sgth}{\gamma}
\newcommand{\gt}{{\tilde g}}

\begin{document}

\title{ Dynamics of a self gravitating light-like matter
shell: a gauge-invariant Lagrangian and Hamiltonian description}

\author{Jacek Jezierski}
\affiliation{Katedra Metod Matematycznych Fizyki, ul. Ho\.{z}a 74,
00-682 Warszawa, Poland}

\author{Jerzy Kijowski}
\affiliation{Centrum Fizyki Teoretycznej PAN, Al.\ Lotnik\'ow
32/46, 02-668 Warszawa, Poland}

\author{Ewa Czuchry}
\affiliation{Katedra Metod Matematycznych Fizyki, ul. Ho\.{z}a 74,
00-682 Warszawa, Poland}


\begin{abstract}
A complete Lagrangian and Hamiltonian description of the theory of
self-gra\-vi\-ta\-ting light-like matter shells is given in terms
of gauge-independent geometric quantities. For this purpose the
notion of an extrinsic curvature for a null-like hypersurface is
discussed and the corresponding Gauss-Codazzi equations are
proved. These equations imply Bianchi identities for spacetimes
with null-like, singular curvature. Energy-momentum tensor-density
of a light-like matter shell is unambiguously defined in terms of
an invariant matter Lagrangian density. Noether identity and
Belinfante-Rosenfeld theorem for such a tensor-density are proved.
Finally, the Hamiltonian dynamics of the interacting system:
``gravity + matter'' is derived from the total Lagrangian, the
latter being an invariant scalar density.
\end{abstract}

\keywords{ general relativity, differential geometry, matter
shells, black holes}

\pacs{04.20.Fy, 04.40.-b, 04.60.Ds}

\maketitle

\section{Introduction}

 Self gravitating matter shell (see \cite{shell-massive,shell1a}) became an important laboratory for testing global
properties of gravitational field interacting with matter. Models
of a thin matter layer allow us to construct useful
mini-superspace examples. Toy models of quantum gravity, based on
these examples may give us a deeper insight into a possible future
shape of the quantum theory of gravity (see \cite{shell-quantum}).
Especially interesting are null-like shells, carrying a
self-gravitating light-like matter (see \cite{hajicek-null}).
Classical equations of motion of such a shell have been derived by
Barrab\`es and Israel in their seminal paper \cite{IB}.

In the present paper we give a complete Lagrangian and Hamiltonian
description of a physical system composed of the gravitational
field interacting with a light-like matter shell. The paper
contains two main results which, in our opinion, improve slightly
the existing classical theory of a null-like shell and provide an
appropriate background for its quantized version. The first result
is the use of fully gauge-invariant, intrinsic geometric objects
encoding physical properties of both the shell (as a null-like
surface in spacetime---see  \cite{nurek}) and the light-like
matter living on the shell. We begin with a description of an
``extrinsic curvature'' of a null-like hypersurface $S$ in terms
of a mixed contravariant-covariant tensor density ${Q}^{a}_{\
b}$---an appropriate null-like analog of the ADM~momentum (cf.
\cite{ADM}). For a non-degenerate (time-like or space-like)
hypersurface, the extrinsic curvature may be described in many
equivalent ways: by tensors or tensor densities, both of them in
the contravariant, covariant or mixed version. In a null-like
case, the degenerate metric on $S$ does not allow us to convert
tensors into tensor densities and {\em vice versa}. Also, we are
not allowed to rise covariant indices, whereas lowering the
contravariant indices is not an invertible operator and leads to
information losses. It turns out that only the mixed tensor
density ${Q}^{a}_{\ b}$ has the appropriate null-like limit and
enables us to formulate the theory of a null-like shell in a full
analogy with the non-degenerate case. We prove Gauss-Codazzi
equations for the extrinsic curvature described by this tensor
density. In particular, the above notion of an extrinsic curvature
may be applied to analyze the structure of non-expanding horizons
(see \cite{7}).

The quantity ${Q}^{a}_{\ b}$ defined in Section \ref{geometry2}
enables us to consider spacetimes with singular
(distribution-like) curvature confined to a null-like
hypersurface, and to prove that the Bianchi identities (understood
in the sense of distributions) are necessary fulfilled in this
case. Such spacetimes are a natural arena for the theory of a
null-like matter shell.

The second main result consists in treating the light-like matter
in a fully dynamical (and not {\em phenomenological}) way. All the
properties of the matter are encoded in a matter Lagrangian, which
is an invariant scalar density on $S$ (no invariant scalar
Lagrangian exists at all for such a matter, because conversion
from scalar densities to scalars and {\em vice versa} is
impossible!). The Lagrangian gives rise to a gauge-invariant
energy-momentum tensor-density ${T^a}_b$, which later---due to
Einstein equations---arises as a source of gravity. Both Noether
and Belinfante-Rosenfeld identities for the quantity ${T^a}_b$ are
proved: they are necessary for the consistence of the theory. We
stress that the contravariant ``symmetric energy-momentum tensor''
$T^{ab}$ cannot be defined unambiguously, whereas the covariant
tensor $T_{ab}$, obtained by lowering the index with the help of a
degenerate metric on $S$, looses partially information contained
in ${T^a}_b$. On the contrary, the mixed contravariant-covariant
tensor density ${T^a}_b$ is unambiguously defined and
contains---as in non-degenerate case---the entire dynamical
information about the underlying matter.

In Section \ref{lagrangian-section} we use a method of variation
of the total (gravity + matter) Lagrangian proposed in
\cite{pieszy} and derive this way Barrab\`es-Israel equations for
gravity, together with the dynamical equations for the matter
degrees of freedom. In Section \ref{hamiltonian-section} we show
how to organize the gravitational and matter degrees of freedom
into a constrained hamiltonian system, with the ADM~mass at
infinity playing the role of the total (``gravity + matter'')
Hamiltonian. Finally, the structure of constraints is analyzed in
Section \ref{constraints-section}. To clarify the exposition of
geometric and physical ideas some of the technical proofs have
been shifted to the Appendix.

\section{Intrinsic geometry of a null hypersurface} \label{geometry1}

A null hypersurface in a Lorentzian space-time $M$ is a
three-dimensional submanifold $S \subset M$ such that the
restriction $g_{ab}$ of the space-time metrics $g_{\mu\nu}$ to $S$
is degenerate.

We shall often use adapted coordinates, where coordinate $x^3$ is
constant on $S$. Space coordinates will be labeled by $k,l =
1,2,3$; coordinates on $S$ will be labeled by $a,b=0,1,2$;
finally, coordinates on $S_{t} := V_{t} \cap S$ (where $V_{t}$ is
a Cauchy surface corresponding to constant value of the
``time-like'' coordinate $x^0=t$) will be labeled by $A,B=1,2$.
Space-time coordinates will be labeled by Greek characters
$\alpha, \beta, \mu, \nu$.

The non-degeneracy of the space-time metric implies that the
metric $g_{ab}$ induced on $S$ from the spacetime metric
$g_{\mu\nu}$ has signature $(0,+,+)$. This means that there is a
non-vanishing null-like vector field $X^a$ on $S$, such that its
four-dimensional embedding $X^\mu$ to $M$ (in adapted coordinates
$X^3=0$) is orthogonal to $S$. Hence, the covector $X_\nu = X^\mu
g_{\mu\nu} = X^a g_{a\nu}$ vanishes on vectors tangent to $S$ and,
therefore, the following identity holds:
\begin{equation}\label{degeneracy}
  X^a g_{ab} \equiv 0 \ .
\end{equation}
It is easy to prove (cf. \cite{JKC}) that integral curves of
$X^a$, after a suitable reparameterization, are geodesic curves of
the space-time metric $g_{\mu\nu}$. Moreover, any null
hypersurface $S$ may always be embedded in a 1-parameter
congruence of null hypersurfaces.

We assume that topologically we have $S = {\mathbb R}^1 \times
S^2$. Since our considerations are purely local, we fix the
orientation of the ${\mathbb R}^1$ component and assume that null
like vectors $X$ describing degeneracy of the metric $g_{ab}$ of
$S$ will be always compatible with this orientation. Moreover, we
shall always use coordinates such that the coordinate $x^0$
increases in the direction of $X$, i.e.,~inequality $X(x^0) = X^0
> 0$ holds. In these coordinates degeneracy fields are of the form
$X = f(\partial_0-n^A\partial_A)$, where $f > 0$, $n_A = g_{0A}$
and we rise indices with the help of the two-dimensional matrix
${\tilde{\tilde g}}^{AB}$, inverse to $g_{AB}$.

If by $\lambda$ we denote the two-dimensional volume form on each
surface $x^0 = \mbox{\rm const.}$:
\begin{equation}\label{lambda}
  \lambda:=\sqrt{\det g_{AB}} \ ,
\end{equation}
then for any degeneracy field $X$ of $g_{ab}$, the following
object
\[
v_{X} := \frac {\lambda}{X(x^0)}
\]
is a scalar density on $S$. Its definition does not depend upon
the coordinate system $(x^a)$ used in the above definition. To
prove this statement is sufficient to show that the value of
$v_{X}$ gets multiplied by the determinant of the Jacobi matrix
when we pass from one coordinate system to another. This means
that ${\bf v}_X := v_X dx^0 \wedge dx^1 \wedge dx^2$ is a
coordinate-independent differential three-form on $S$. However,
$v_X$ depends upon the choice of the field  $X$.

It follows immediately from the above definition that the
following object:
\[
\Lambda = v_X \ X \ ,
\]
is a well defined (i.e.,~coordinate-independent) vector-density on
$S$. Obviously, it {\em does not depend} upon any choice of the
field $X$:
\begin{equation}\label{Lambda}
\Lambda =  \lambda (\partial_0-n^A\partial_A) \ .
\end{equation}
Hence, it is an intrinsic property of the internal geometry
$g_{ab}$ of $S$. The same is true for the divergence $\partial_a
\Lambda^a$, which is, therefore, an invariant, $X$-independent,
scalar density on $S$. Mathematically (in terms of differential
forms), the quantity $\Lambda$ represents the two-form:
\[
{\bf L} := \Lambda^a \left( \partial_a \  \rfloor \  dx^0 \wedge
dx^1 \wedge dx^2 \right) \ ,
\]
whereas the divergence represents its exterior derivative (a
three-from): d${\bf L} := \left( \partial_a \Lambda^a \right)dx^0
\wedge dx^1 \wedge dx^2$. In particular, a null surface with
vanishing d${\bf L}$ is called a {\em non-expanding horizon} (see
\cite{7}).

Both objects ${\bf L}$ and ${\bf v}_X$ may be defined
geometrically, without any use of coordinates. For this purpose we
note that at each point $x \in S$, the tangent space $T_xS$ may be
quotiented with respect to the degeneracy subspace spanned by $X$.
The quotient space carries a non-degenerate Riemannian metric and,
therefore, is equipped with a volume form $\omega$ (its coordinate
expression would be: $\omega = \lambda \ dx^1 \wedge dx^2$). The
two-form ${\bf L}$ is equal to the pull-back of $\omega$ from the
quotient space to $T_xS$. The three-form ${\bf v}_X$ may be
defined as a product: ${\bf v}_X = \alpha \wedge {\bf L}$, where
$\alpha$ is {\em any} one-form on $S$, such that $<X,\alpha>
\equiv 1$.

The degenerate metric $g_{ab}$ on $S$ does not allow to define
{\em via} the compatibility condition $\nabla g = 0$, any natural
connection, which could apply to generic tensor fields on $S$.
Nevertheless, there is one exception: we are going to show that
the degenerate metric defines {\em uniquely} a certain covariant,
first order differential operator which will be extensively used
in our paper. The operator may be applied only to mixed
(contravariant-covariant) tensor-density fields ${\bf H}^{a}_{\
b}$, satisfying the following algebraic identities:
\begin{eqnarray}
{\bf H}^{a}_{\ b} X^b = 0 \ , \label{G-1} \\ {\bf H}_{ab} = {\bf
H}_{ba} \ , \label{G-2}
\end{eqnarray}
where ${\bf H}_{ab} := g_{ac} {\bf H}^{c}_{\ b}$. Its definition
cannot be extended to other tensorial fields on $S$. Fortunately,
as will be seen, extrinsic curvature of a null-like surface and
the energy-momentum tensor of a null-like shell are described by
tensor-densities of this type.

The operator, which we denote by ${\overline{\nabla} }_a {\bf
H}^{a}_{\ b}$, could be defined by means of the four dimensional
metric connection in the ambient space-time $M$ in the following
way. Given ${\bf H}^{a}_{\ b}$, take any its extension ${\bf
H}^{\mu\nu}$  to a four-dimensional, symmetric tensor density,
``orthogonal'' to $S$, i.e. satisfying ${\bf H}^{\perp\nu}=0$
(``$\perp$'' denotes the component transversal to $S$). Define
${\overline{\nabla} }_a {\bf H}^{a}_{\ b}$ as the restriction to
$S$ of the four-dimensional covariant divergence ${\nabla}_\mu
{\bf H}^{\mu}_{\ \nu}$. As will be seen in the sequel, ambiguities
which arise when extending three dimensional object ${\bf
H}^{a}_{\ b}$ living on $S$ to the four dimensional one, cancel
finally and the result is unambiguously defined as a
covector-density on $S$. It turns out, however, that this result
does not depend upon the space-time geometry and may be defined
intrinsically on $S$. This is why we first give this intrinsic
definition, in terms of the degenerate metric.

In case of a non-degenerate metric, the covariant divergence of a
symmetric tensor ${\bf H}$ density may be calculated by the
following formula:
\begin{eqnarray}
    \nonumber
    {\nabla}_a {\bf H}^{a}_{\ b} & = &
    \partial_a {\bf H}^{a}_{\ b} - {\bf H}^{a}_{\ c}
    {\Gamma}^c_{ab} \\ \label{covariant-deg}
    & = & \partial_a {\bf H}^{a}_{\ b} -
    \frac 12 {\bf H}^{ac} g_{ac , b} \ ,
\end{eqnarray}
with $g_{ac , b} := \partial_b g_{ac} $. In case of our degenerate
metric, we want to mimic the last formula, but here rising of
indices of  ${\bf H}^{a}_{\ b}$ makes no sense. Nevertheless,
formula (\ref{covariant-deg}) may be given a unique sense also in
the degenerate case, if applied to a tensor density ${\bf
H}^{a}_{\ b}$ satisfying identities (\ref{G-1}) and (\ref{G-2}).
Namely, we take as ${\bf H}^{ac}$ {\em any} symmetric tensor
density, which reproduces ${\bf H}^{a}_{\ b}$ when lowering an
index:
\begin{equation}\label{G-mixed}
  {\bf H}^{a}_{\ b}  = {\bf H}^{ac} g_{cb} \ .
\end{equation}
It is easily seen, that such a tensor-density always exists due to
identities (\ref{G-1}) and (\ref{G-2}), but reconstruction of
${\bf H}^{ac}$ from ${\bf H}^{a}_{\ b}$ is not unique, because
${\bf H}^{ac} + C X^a X^c$ also satisfies ({\ref{G-mixed}) if
${\bf H}^{ac}$ does. Conversely, two such symmetric tensors ${\bf
H}^{ac}$ satisfying (\ref{G-mixed}) may differ only by $C X^a
X^c$. This non-uniqueness does not influence the value of
(\ref{covariant-deg}), because of the following identity implied
by (\ref{degeneracy}):
\begin{align}\label{identi}
  0 &\equiv (X^a X^c g_{ac} )_{,b} \nonumber\\&= X^a X^c g_{ac,b} + 2 X^a g_{ac}
X^c{_{,b}} =  X^a X^c g_{ac,b} \, .
\end{align}
Hence, the following definition makes sense:
\begin{equation}\label{div-final}
  {\overline{\nabla} }_a {\bf H}^{a}_{\ b}  :=
\partial_a {\bf H}^{a}_{\ b} - \frac 12 {\bf H}^{ac}
g_{ac , b} \ .
\end{equation}
The right-hand-side does not depend upon any choice of coordinates
(i.e.,~transforms like a genuine covector-density under change of
coordinates). The proof is straightforward and does not differ
from the standard case of formula (\ref{covariant-deg}), when
metric $g_{ab}$ is non-degenerate.

To express directly the result in terms of the original tensor
density ${\bf H}^{a}_{\ b}$, we observe that it has five
independent components and may be uniquely reconstructed from
${\bf H}^{0}_{\ A}$ (2 independent components) and the symmetric
two-dimensional matrix ${\bf H}_{AB}$ (3 independent components).
Indeed, identities (\ref{G-1}) and (\ref{G-2}) may be rewritten as
follows:
\begin{align}
{\bf H}^{A}_{\ B} & =  {\tilde{\tilde g}}^{AC}{\bf H}_{CB} - n^A
{\bf H}^{0}_{\ B} \ , \label{AB}
\\ {\bf H}^{0}_{\ 0} & =  {\bf H}^{0}_{\ A} n^A \ ,  \label{00}
\\ {\bf H}^{B}_{\ 0} & =  \left( {\tilde{\tilde g}}^{BC}{\bf H}_{CA}
- n^B {\bf H}^{0}_{\ A} \right) n^A \label{B0} \ .
\end{align}
The correspondence between ${\bf H}^{a}{_b}$ and $({\bf
H}^{0}{_A},{\bf H}_{AB})$ is one-to-one.

To reconstruct ${\bf H}^{ab}$  from ${\bf H}^{a}{_b}$ up to an
arbitrary additive term $C X^a X^b$, take the following,
coordinate dependent, symmetric quantity:
\begin{align}
{\bf F}^{AB} & :=  {\tilde{\tilde g}}^{AC} {\bf H}_{CD }
{\tilde{\tilde g}}^{DB} - n^A {\bf H}^{0}_{\ C } {\tilde{\tilde
g}}^{CB} - n^B {\bf H}^{0}_{\ C } {\tilde{\tilde g}}^{CA} \ ,
\\ {\bf F}^{0A} & :=  {\bf H}^{0}_{\ C } {\tilde{\tilde g}}^{CA}
=: {\bf F}^{A0} \ ,
\\ {\bf F}^{00} & :=  0 \ .
\end{align}
It is easy to observe that any ${\bf H}^{ab}$ satisfying
(\ref{G-mixed}) must be of the form:
\begin{equation}\label{reconstr}
{\bf H}^{ab} = {\bf F}^{ab} + {\bf H}^{00} X^a X^b \ .
\end{equation}
The non-uniqueness in the reconstruction of ${\bf H}^{ab}$ is,
therefore, completely described by the arbitrariness in the choice
of the value of ${\bf H}^{00}$. Using these results we finally
obtain:
\begin{eqnarray}
{\overline{\nabla} }_a {\bf H}^{a}_{\ b} & := &
\partial_a {\bf H}^{a}_{\ b} - \frac 12 {\bf H}^{ac}
g_{ac , b} =
\partial_a {\bf H}^{a}_{\ b} - \frac 12 {\bf F}^{ac}
g_{ac , b}\nonumber
\\  & = & \partial_a {\bf H}^{a}_{\ b} - \frac 12 \left(
2 {\bf H}^{0}_{\ A} \ n^A_{ \ ,b} - {\bf H}_{AC}  {\tilde{\tilde
g}}^{AC}_{\ \ \ ,b} \right) \ . \label{Coda}
\end{eqnarray}
The operator on the right-hand side of (\ref{Coda}) may thus be
called the (three-dimen\-sio\-nal) covariant derivative of ${\bf
H}^{a}_{\ b}$ on $S$ with respect to its degenerate metric
$g_{ab}$. We have just proved that it is well defined
(i.e.,~coordinate-independent) for a tensor density ${\bf
H}^{a}_{\ b}$ fulfilling conditions (\ref{G-1}) and (\ref{G-2}).

Equation (\ref{div-final}) suggests yet another definition of the
covariant divergence operator. At a given point $x \in S$ choose
any coordinate system, such that derivatives of the metric
components $g_{ac}$ vanish at $x$, i.e.,~$g_{ac , b}(x)=0$. Such a
coordinate system may be called {\em inertial}. The covariant
divergence may thus be defined as a partial divergence but
calculated in an inertial system: ${\overline{\nabla} }_a {\bf
H}^{a}_{\ b} := \partial_a {\bf H}^{a}_{\ b}$. Ambiguities in the
choice of an inertial system do not allow us to extend this
definition to a genuine covariant derivative ${\overline{\nabla}
}_c {\bf H}^{a}_{\ b}$. However, it may be easily checked that
they are sufficiently mild for an unambiguous definition of the
divergence (cf. Remark at the end of Section
\ref{energy-momentum-tensor}).

The above two equivalent definitions of the operator
${\overline{\nabla}}$ use only the intrinsic metric of $S$. We
want to prove now that they coincide with the definition given in
terms of the four dimensional space-time metric-connection. For
that purpose observe, that the only non-uniqueness in the
reconstruction of the four-dimensional tensor density of ${\bf
H}^{\mu\nu}$ is of the type $C X^\mu X^\nu$. Indeed, any such
reconstruction may be obtained from a reconstruction of ${\bf
H}^{ac}$ by putting ${\bf H}^{3\nu} = 0$ in a coordinate system
adapted to $S$ (i.e.,~such that the coordinate $x^3$ remains
constant on $S$). Now, calculate the four-dimensional covariant
divergence ${\bf H}_\nu := \nabla_\mu {\bf H}^{\mu}_{\ \nu}$. Due
to the geodesic character of integral curves of the field $X$, the
only non-uniqueness which remains after this operation is of the
type ${\tilde C} X_\nu$. Hence, the restriction ${\bf H}_b$ of
${\bf H}_\nu$ to $S$ is already unique. Due to
(\ref{covariant-deg}), it equals:
\begin{eqnarray}\label{int-ext}
  \nabla_\mu {\bf H}^{\mu}_{\ b} & = &
  \partial_\mu {\bf H}^{\mu}_{\ b} -
    \frac 12 {\bf H}^{\mu\lambda} g_{\mu\lambda , b}
    \nonumber \\
    & = &
    \partial_a {\bf H}^{a}_{\ b} -
    \frac 12 {\bf H}^{ac} g_{ac , b} =
    {\overline{\nabla} }_a {\bf H}^{a}_{\ b}\ .
\end{eqnarray}

\section{Extrinsic geometry of a null hypersurface.
Gauss-Codazzi equations} \label{geometry2}

To describe exterior geometry of $S$ we begin with covariant
derivatives {\em along} $S$ of the ``orthogonal vector $X$''.
Consider the tensor $\nabla_a X^\mu$. Unlike in the non-degenerate
case, there is no unique ``normalization'' of $X$ and, therefore,
such an object does depend upon a choice of the field $X$. The
length of $X$ is constant (because vanishes). Hence, the tensor is
again orthogonal to $S$, i.e.,~the components corresponding to
$\mu = 3$ vanish identically in adapted coordinates. This means
that $\nabla_a X^b$ is a purely three dimensional tensor living on
$S$. For our purposes it is useful to use the ``ADM-like'' version
of this object, defined in the following way:
\begin{equation}\label{Q-fund}
{Q^a}_b (X) := -s \left\{ v_X \left( \nabla_b X^a - \delta_b^a
\nabla_c X^c \right) + \delta_b^a \partial_c \Lambda^c \right\} \
,
\end{equation}
where $s:=\sgn g^{03}=\pm 1$. Due to above convention, the
``extrinsic curvature'' ${Q^a}_b (X)$ feels only {\em external
orientation} of $S$ and does not feel any internal orientation of
the field $X$.

{\bf Remark:} If $S$ is a {\em non-expanding horizon}, the last
term in the above definition vanishes.

The last term in (\ref{Q-fund}) is $X$-independent. It has been
introduced in order to correct algebraic properties of the
quantity $ v_X \left( \nabla_b X^a - \delta_b^a \nabla_c X^c
\right)$: we prove in the Appendix \ref{GuuA} (see Remark after
(\ref{Qli})) that ${Q^a}_b$ satisfies identities
(\ref{G-1})--(\ref{G-2}) and, therefore, its covariant divergence
with respect to the degenerate metric $g_{ab}$ on $S$ is uniquely
defined. This divergence enters into the Gauss-Codazzi equations
which we are going to formulate now. Gauss-Codazzi equations
relate the divergence of $Q$ with the transversal component ${\cal
G}^{\perp}_{\ b}$ of the Einstein tensor-density ${\cal G}^\mu_{\
\nu} = \sqrt{|\det g |} \left( R^\mu_{\ \nu} - \delta^\mu_\nu
\frac 12 R \right)$. The transversal component of such a
tensor-density is a well defined three-dimensional object living
on $S$. In coordinate system adapted to $S$, i.e.,~such that the
coordinate $x^3$ is constant on $S$, we have ${\cal G}^{\perp}_{\
b} = {\cal G}^{3}_{\ b}$. Due to the fact that ${\cal G}$ is a
tensor-density, components ${\cal G}^{3}_{\ b}$ {\em do not
change} with changes of the coordinate $x^3$, provided it remains
constant on $S$. These components describe, therefore, an
intrinsic covector-density living on $S$.

\begin{theo}
The following null-like-surface version of the Gauss-Codazzi
equation is true:
\begin{equation}\label{G-C}
    {\overline{\nabla} }_a {Q}^{a}_{\ b}(X) +s v_{X} \partial_b
    \left( \frac {\partial_c \Lambda^c}{v_{X}} \right) \equiv
    -{\cal G}^{\perp}_{\ b}  \ .
\end{equation}
 \end{theo}
We remind the reader that the ratio between two scalar densities:
$\partial_c \Lambda^c$ and $v_X$, is a scalar function. Its
gradient is a co-vector field. Finally, multiplied by the density
$v_X$, it produces an intrinsic co-vector density on $S$. This
proves that also the left-hand-side is a well defined, geometric
object living on $S$.

To prove consistency of (\ref{G-C}), we must show that the
left-hand side does not depend upon a choice of $X$. For this
purpose consider another degeneracy field: $fX$, where $f > 0$ is
a function on $S$. We have:
\begin{align}\label{gauge}
  -s{Q}^{a}_{\ b}(fX) & =  v_{fX} \left( \nabla_b (f  X^a ) -
  \delta_b^a \nabla_c ( f X^c )
\right) + \delta_b^a \partial_c \Lambda^c \nonumber
\\ & =  \frac
1f v_{X} \left( f \nabla_b  X^a  +X^a \partial_b f -
  \delta_b^a f \nabla_c  X^c \right.\nonumber\\
  & \left. \qquad- \delta_b^a X^c \partial_c f
\right) + \delta_b^a \partial_c \Lambda^c \nonumber
\\  & =  -s{Q}^{a}_{\ b}(X) + \Lambda^a \varphi_{,b} -
\delta_b^a \Lambda^c \varphi_{,c} \ ,
\end{align}
 where $\varphi := \log f$. It is easy to see that the
tensor
\begin{equation}\label{q}
  q^{a}_{\ b}(\varphi):= \Lambda^a \varphi_{,b} -
\delta_b^a \Lambda^c \varphi_{,c} \ ,
\end{equation}
satisfies identity (\ref{G-1}). Moreover, $q_{ab} = - g_{ab}
\Lambda^c \varphi_{,c}$, which proves (\ref{G-2}). On the other
hand, we have
\begin{equation}\label{dodatek}
   v_{fX} \partial_b \left(
    \frac {\partial_c \Lambda^c}{v_{fX}} \right) =  v_{X} \partial_b
    \left(
    \frac {\partial_c \Lambda^c}{v_{X}}\right) + (\partial_c \Lambda^c)
    \varphi_{,b} \ ,
\end{equation}
But, using formula (\ref{Coda}) we immediately get:
\[
{\overline{\nabla} }_a {q}^{a}_{\ b}(\varphi) = (\partial_c
\Lambda^c)
    \varphi_{,b} \ ,
    \]
which proves that the left hand side of (\ref{G-C}) does not
depend upon any choice of the field $X$. The complete proof of the
Gauss-Codazzi equation (\ref{G-C}) is given in the Appendix
\ref{GuuA}\footnote{In non-degenerate case, there are four
independent Gauss-Codazzi equations: besides ${\cal G}^{\perp}_{\
b}$, there is an additional equation relating ${\cal G}^{\perp}_{\
\perp}$ with (external and internal) geometry of $S$. In
degenerate case, vector orthogonal to $S$ is---at the same
time---tangent to it. Hence, ${\cal G}^{\perp}_{\ \perp}$ is a
combination of quantities ${\cal G}^{\perp}_{\ b}$ and there are
only three independent Gauss-Codazzi equations.}.

\section{Bianchi identities for space-times with distribution
valued curvature} \label{Bianchi}

In this paper we consider a space-time $M$ with distribution
valued curvature tensor in the sense of Taub \cite{Taub}. This
means that the metric tensor, although continuous, is not
necessarily $C^1$-smooth across $S$: we assume that the connection
coefficients $\Gamma^\lambda_{\mu\nu}$ may have only step
discontinuities (jumps) across $S$. Formally, we may calculate the
Riemann curvature tensor of such a spacetime, but derivatives of
these discontinuities with respect to the variable $x^3$ produce a
$\boldsymbol\delta$-like, singular part of $R$:
\begin{equation}\label{R-sing}
  \mbox{\rm sing}{(R)^\lambda}_{\mu\nu\kappa} = \left( \delta^3_\nu
[ \Gamma^\lambda_{\mu\kappa} ] - \delta^3_\kappa [
\Gamma^\lambda_{\mu\nu} ] \right) \boldsymbol\delta (x^3) \ ,
\end{equation}
where by $\boldsymbol\delta$ we denote the Dirac distribution (in
order to distinguish it from the Kronecker symbol $\delta$) and by
$[f]$ we denote the jump of a discontinuous quantity $f$ between
the two sides of $S$. Above formula is invariant under {\em
smooth} transformations of coordinates. There is, however, no
sense to impose such a smoothness across $S$. In fact, the
smoothness of spacetime is an independent condition on both sides
of $S$. The only reasonable assumption imposed on the
differentiable structure of $M$ is that the metric tensor---which
is smooth separately on both sides of $S$---remains
continuous\footnote{Many authors insist in relaxing this condition
and assuming only the continuity of the three-dimensional
intrinsic metric on $S$. We stress that the (apparently stronger)
continuity condition for the four-dimensional metric does not lead
to any loss of generality and may be treated as an additional,
technical gauge imposed {\em not upon the physical system} but
upon its mathematical parameterization. We discuss thoroughly this
issue in a Remark at the end of the present Section.} across $S$.
Admitting coordinate transformations preserving the above
condition, we loose a  part of information contained in quantity
(\ref{R-sing}), which becomes now coordinate-dependent. It turns
out, however, that another part, namely the Einstein
tensor-density calculated from (\ref{R-sing}), preserves its
geometric, intrinsic (i.e.,~coordinate-independent) meaning. In
case of a non-degenerate geometry of $S$, the following formula
was used by many authors (see \cite{shell-massive, shell1a,
shell-quantum,shell1,shell1b}):
\begin{equation}\label{E-sing-non-deg}
\mbox{\rm sing}({\cal G})^{\mu\nu }  =   {\bf G}^{\mu\nu}
\boldsymbol\delta(x^3) \ ,
\end{equation}
where the ``transversal-to-$S$'' part of ${\bf G}^{\mu\nu}$
vanishes identically:
\begin{equation}\label{G^perp=0}
{\bf G}^{\perp \nu} \equiv 0 \ ,
\end{equation}
and the ``tangent-to-$S$'' part ${\bf G}^{ab}$ equals to the jump
of the ADM~extrinsic curvature $Q^{ab}$ of $S$ between the two
sides of the surface:
\begin{equation}\label{G-Q}
  {\bf G}^{ab} = [Q^{ab}] \ .
\end{equation}
This quantity is a purely {\em three-dimensional}, symmetric
tensor-density living on $S$. When multiplied by the {\em
one-dimensional} density $\boldsymbol\delta(x^3)$ in the
transversal direction, it produces the {\em four-dimensional}
tensor density ${\cal G}$ according to formula
(\ref{E-sing-non-deg}).

Now, let us come back to the case of our degenerate surface $S$.
One of the goals of the present paper is to prove, that formulae
(\ref{E-sing-non-deg}) and (\ref{G^perp=0}) remain valid also in
this case. In particular, the latter formula means that the
four-dimensional quantity ${\cal G}^{\mu\nu }$ reduces in fact to
an intrinsic, three-dimensional quantity living on $S$. However,
formula (\ref{G-Q}) cannot be true, because---as we have
seen---there is no way to define uniquely the object $Q^{ab}$ for
the degenerate metric on $S$. Instead, we are able to prove the
following formula:
\begin{equation}\label{Einst-Q}
  {{\bf G}^a}_b = [ {Q}^{a}_{\ b}(X) ] \ ,
\end{equation}
where the bracket denotes the jump of ${Q}^{a}_{\ b}(X)$ between
the two sides of the singular surface. Observe that this quantity
{\em does not depend} upon any choice of $X$. Indeed, formula
(\ref{gauge}) shows that $Q$ changes identically on both sides of
$S$ when we change $X$ and, hence, these changes cancel. This
proves that the singular part $\mbox{\rm sing}{({\cal G})^a}_b$ of
the Einstein tensor is well defined.

{\bf Remark:} Otherwise as in the non-degenerate case, the
contravariant components ${\bf G}^{ab}$ in formula
(\ref{E-sing-non-deg}) do not transform as a tensor-density on
$S$. Hence, the quantity defined by these components would be
coordinate-dependent. According to (\ref{Einst-Q}), ${\bf G}$
becomes an intrinsic 3-dimensional tensor-density on $S$ only
after lowering an index, i.e.,~in the version of ${{\bf G}^a}_b$.
This proves that ${\bf G}^{\mu\nu}$ may be reconstructed from
${{\bf G}^a}_b$ up to an additive term $C X^\mu X^\nu$ only. We
stress that the dynamics of the shell, which we discuss in the
sequel, is unambiguously expressed in terms of the
gauge-invariant, intrinsic quantity ${{\bf G}^a}_b$.

Proofs of the above facts are given in the Appendix \ref{GuuA}.

We conclude that the total Einstein tensor of our spacetime is a
sum of the regular part\footnote{The regular part is a smooth
tensor density on both sides of the surface $S$ (calculated for
the metric $g$ separately) with possible step discontinuity across
$S$.} $\mbox{\rm reg}({\cal G})$ and the above singular part
$\mbox{\rm sing}({\cal G})$ living on the singularity surface $S$.
Thus
\begin{equation}\label{Bianchi-1}
  {{\cal G}^\mu}_\nu = \mbox{\rm reg}{({\cal G})^\mu}_\nu +
  \mbox{\rm sing}{({\cal G})^\mu}_\nu \ ,
\end{equation}
and the singular part is given {\em up to an additive term} $C
X^\mu X_\nu \boldsymbol\delta(x^3)$. Due to (\ref{identi}), the
following {\em four-dimensional} covariant divergence is
unambiguously defined:
\begin{equation}
0 = \nabla_\mu {\cal G}^{\mu}_{\ c} =
\partial_\mu {\cal G}^{\mu}_{\ c}
- {\cal G}^{\mu}_{\ \alpha} \Gamma^\alpha_{\mu c} = \partial_\mu
{\cal G}^{\mu}_{\ c} - \frac 12 {\cal G}^{\mu\lambda}
g_{\mu\lambda , c} \ .\label{bian-ident}
\end{equation}
We are going to prove that this quantity vanishes identically.
Indeed, the regular part of this divergence vanishes on both sides
of $S$ due to Bianchi identities: $\mbox{\rm reg}\left( \nabla_\mu
{\cal G}^{\mu}_{\ c} \right)\equiv 0$. As a next step we observe
that the singular part is proportional to
$\boldsymbol\delta(x^3)$, i.e.,~that the Dirac delta contained in
$\mbox{\rm sing}({\cal G})$ will not be differentiated, when we
apply the above covariant derivative to the singular part
(\ref{E-sing-non-deg}). This is true because $\mbox{\rm
sing}({\cal G})^{3}_{\ \nu} = 0$. Hence, only the covariant
divergence of ${\bf G}$ along $S$ (multiplied by
$\boldsymbol\delta(x^3)$) remains. Another
$\boldsymbol\delta$-like term is obtained from $\partial_\mu {\cal
G}^{\mu}_{\ c}$, when applied to the (piecewise continuous)
regular part of ${\cal G}$. This way we obtain the term
$[\mbox{\rm reg}({\cal G})^{\perp}_{\ c}] \boldsymbol\delta(x^3)
$. Finally, the total singular part of the Bianchi identities
reads:
\begin{equation}\label{Bianchi-fund}
    \mbox{\rm sing}\left(  \nabla_\mu {\cal G}^{\mu}_{\ c} \right)
     = \left(
  [\mbox{\rm reg}({\cal G})^{\perp}_{\ c}] +
  {\overline{\nabla} }_a {\bf G}^{a}_{\ b} \right)
  \boldsymbol\delta(x^3) \equiv 0
    \ ,
\end{equation}
and vanishes identically due to the Gauss-Codazzi equation
(\ref{G-C}), when we calculate its jump across $S$. Hence, we have
proved that the Bianchi identity $\nabla_\mu {\cal G}^{\mu}_{\ c}
\equiv 0$ holds universally (in the sense of distributions) for
spacetimes with singular, light-like curvature.

It is worthwhile to notice that the last term in definition
(\ref{Q-fund}) of the tensor-density $Q$ of $S$ is identical on
its both sides. Hence, its jump across $S$ vanishes identically.
This way the singular part of the Einstein tensor density
(\ref{Einst-Q}) reduces to:
\begin{equation}\label{Einst-Q-1}
  {{\bf G}^a}_b = [ {Q}^{a}_{\ b} ] = -s
   v_X  \left( [\nabla_b X^a ] - \delta_b^a [\nabla_c X^c ]\right)
     \ .
\end{equation}

{\bf Remark:} Possibility of defining the singular Einstein tensor
and its divergence {\em via} the standard formulae of Riemannian
geometry (but understood in the sense of distribution!) simplifies
considerably the mathematical description of the theory. This
techniques is based, however, on the continuity assumption for the
four-dimensional metric. This is not a geometric or physical
condition imposed on the system, but only the coordinate (gauge)
condition. Indeed, whenever the three-dimensional, internal metric
on $S$ is continuous, also the remaining four components of the
total metric can be made continuous by a simple change of
coordinates. In this new coordinate system we may use our
techniques based on the theory of distributions and derive both
the Lagrangian and the Hamiltonian version of the dynamics of the
total (``gravity + shell'') system. As will be seen in the sequel,
the dynamics derived this way does not depend upon our gauge
condition and is expressed in terms of equations which make sense
also in general coordinates. As an example of such an equation
consider (\ref{Einst-Q}) which---even if derived here by technique
of distributions under more restrictive conditions---remains valid
universally. We stress that even in a smooth, vacuum spacetime (no
shell at all!) one can consider non-smooth coordinates, for which
only the internal metric $g_{ab}$ on a given surface, say $\{x^3 =
C\}$, is continuous, whereas the remaining four components $g_{3
\mu}$ may have jumps. The entire Canonical Gravity may be
formulated in these coordinates. In particular, the Cauchy
surfaces $\{x^0 =\, $const.$\}$ would be allowed to be non-smooth
here. Nobody uses such a formulation (even if it is fully
legitime) because of its relative complexity: the additional gauge
condition imposing the continuity of the whole four-dimensional
metric makes life much easier!

\section{Energy-momentum tensor of a light-like matter.
Belinfante-Rosenfeld identity} \label{energy-momentum-tensor}

The goal of this paper is to describe  interaction between a thin
light-like matter-shell and the gravitational field. We derive all
the properties of such a matter from its Lagrangian density ${
L}$. It may depend upon (non-specified) matter fields $z^K$ living
on a null-like surface $S$, together with their first derivatives
${z^K}_a:= \partial_a z^K$ and---of course---the (degenerate)
metric tensor $g_{ab}$ of $S$:
\begin{equation}\label{L1}
L=L(z^K;{z^K}_a;  g_{ab}) \ .
\end{equation}
We assume that $L$ is an invariant scalar density on $S$.
Similarly as in the standard case of canonical field theory,
invariance of the Lagrangian with respect to reparametrizations of
$S$ implies important properties of the theory: the
Belinfante-Rosenfeld identity and the Noether theorem, which will
be discussed in this Section. To get rid of some technicalities,
we assume in this paper that the matter fields $z^K$ are
``spacetime scalars'', like, e.g.,~material variables of any
thermo-mechanical theory of continuous media (see, e.g.,
 \cite{shell1,KSG}). This means that the Lie derivative
${\cal L}_Y z$ of these fields with respect to a vector field $Y$
on $S$ coincides with the partial derivative:
\[
({\cal L}_Y z)^K = {z^K}_a \ Y^a \ .
\]
The following Lemma characterizes Lagrangians which fulfill the
invariance condition:
\begin{lemma}\label{lem1}
Lagrangian density (\ref{L1}) concentrated on a null hypersurface
$S$ is invariant if and only if it is of the form:
\begin{equation}\label{lagr-form}
L=v_{X} f(z ; {\cal L}_X z ; g)\ ,
\end{equation}
where $X$ is any degeneracy field of the metric $g_{ab}$ on $S$
and $f(\cdot \ ; \cdot \ ; \cdot)$ is a scalar function,
homogeneous of degree 1 with respect to its second variable.
\end{lemma}
Proof of the Lemma and examples of invariant Lagrangians for
different light-like matter fields are given in Appendix
\ref{proof-matter-Lagrangian}.

{\bf Remark:} Because of the homogeneity of $f$ with respect to
${\cal L}_X z$, the above quantity does not depend upon a choice
of the degeneracy field $X$.

Dynamical properties of such a matter are described by its
canonical energy-momentum tensor-density, defined in a standard
way:
\begin{equation}\label{e-m-can}
  {T^a}_b := \frac{\partial { L}}{\partial {z^K}_a}
  {z^K}_b - \delta^a_{\; b} { L} \ .
\end{equation}
It is ``symmetric'' in the following sense:
\begin{theo}\label{th1}
Canonical energy-momentum tensor-density ${T^a}_b$ constructed
from an invariant Lagrangian density fulfills identities
(\ref{G-1}) and (\ref{G-2}), i.e.,~the following holds:
\begin{equation}
  {T^a}_b X^b=0\;\;\; {\rm and}\;\;\; T_{ab}=T_{ba}\ .
\end{equation}
\end{theo}
\begin{proof}[Proof:]
For a Lagrangian density of the form (\ref{lagr-form}) we have:
\begin{align}
{T^a}_b&=\frac{\partial L}{\partial {z^K}_a} {z^K}_b -
\delta^a_{\; b} L\nonumber\\ & =v_{X}  \left( X^a \frac{\partial
f}{\partial ({z^K}_d \ X^d)} {z^K}_b-\delta^a_{\; b} f \right) \ ,
\end{align}
whence:
\begin{equation}
T_{ab}={T^c}_b g_{ca}=-v_{X} f g_{ab}=T_{ba} \ .
\end{equation}
Homogeneity of $f$ with respect to the argument $({z^K}_d \ X^d)$
implies:
\begin{equation}
T^a{_b} X^b=v_{X} X^a  \left(\frac{\partial f}{\partial({z^K}_d \
X^d)} \left({z^K}_b \ X^b \right) - f \right) = 0\ .
\end{equation}

\end{proof}

In case of a non-degenerate geometry of $S$, one considers also
the ``symmetric energy-momentum tensor-density'' ${\tau}^{ab}$,
defined as follows:
\begin{equation}\label{seym}
{\tau}^{ab}:=2 \frac{\partial L}{\partial g_{ab}} \ .
\end{equation}
In our case the degenerate metric fulfills the constraint: $\det
g_{ab} \equiv 0$. Hence, the above quantity {\em is not} uniquely
defined. However, we may define it, but only {\em up to an
additive term} equal to the annihilator of this constraint. It is
easy to see that the annihilator is of the form $C X^a X^b$.
Hence, ambiguity in the definition of the symmetric
energy-momentum tensor is precisely equal to ambiguity in the
definition of $T^{ab}$, if we want to reconstruct it from the well
defined object ${T^a}_b$ . This ambiguity is cancelled when we
lower an index. We shall prove in the next theorem, that for field
configurations satisfying field equations, both the canonical and
the symmetric tensors coincide\footnote{In our convention, energy
is described by formula: $H={T^0}_0 = {p_K}^0 \dot{z}^K - L \ge
0$, analogous to $H=p \dot{q}-L$ in mechanics and well adapted for
Hamiltonian purposes. This convention differs from the one used in
\cite{Misner}, where energy is given by $T_{00}$. To keep standard
conventions for Einstein equations, we take standard definition of
the {\em symmetric} energy-momentum tensor ${\tau^a}_b$. This is
why Belinfante-Rosenfeld theorem takes form
${\tau^a}_b=-{T^a}_b$.}. This is an analog of the standard
Belinfante-Rosenfeld identity (see \cite{R-B}). Moreover, Noether
theorem (vanishing of the divergence of $T$) is true. We summarize
these facts in the following:

\begin{theo}\label{ros-bel}
If $L$ is an invariant Lagrangian and if the field configuration
$z^K$ satisfies Euler-Lagrange equations derived from $L$:
\begin{equation}\label{E-L}
  \frac{\partial { L}}{\partial z^K}-\partial_a \frac{\partial
{ L}}{\partial {z^K}_a} = 0 \ ,
\end{equation}
then the following statements are true:
\begin{enumerate}
  \item Belinfante-Rosenfeld identity: canonical
  energy-momentum tensor ${T^a}_b$ coincides with
  (minus---because of the convention used)
  symmetric energy-mo\-men\-tum tensor ${\tau}^{ab}$:
\begin{equation}\label{R-B}
 {T^a}_b = - {\tau}^{ac}g_{cb} \ ,
\end{equation}
  \item Noether Theorem:
  \begin{equation}\label{Noether}
  {\overline{\nabla} }_a {T^a}_b =0\ .
\end{equation}
\end{enumerate}

\end{theo}

\begin{proof}[Proof:]
Invariance of the Lagrangian with respect to space-time
diffeomorphisms generated by a vector field $Y$ on $S$ means that
transporting the arguments $(z;\partial z;  g)$ of $L$ along $Y$
gives the same result as transporting directly the value of the
scalar density $L$ on $S$:
\begin{equation}
\frac{\partial { L}}{\partial z^K}({\cal L}_Y z)^K+ \frac{\partial
{ L}}{\partial {z^K}_a}{({\cal L}_Y z)^K}_a +\frac{\partial {
L}}{\partial g_{ac}}({\cal L}_Y g)_{ac}={\cal L}_Y
 L \ .
\end{equation}
Take for simplicity $Y=\frac{\partial }{\partial x^b}$ (or $Y^a =
\delta^a_{\; b}$). Hence, we have: ${({\cal L}_Y z)^K}_a =
{z^K}_{ba} = {z^K}_{ab}$. Applying this and rearranging terms in
the above expression we obtain:
\begin{align}\label{expr10}
&\left(\frac{\partial { L}}{\partial z^K}-\partial_a
\frac{\partial { L}}{\partial {{z^K}_a}}\right)
{{z^K}_b}\nonumber\\ &\qquad+\partial_a\left(\frac{\partial {
L}}{\partial {{z^K}_a}} {{z^K}_b}- \delta^a_{\; b} {
L}\right)+\frac{\partial L}{\partial g_{ ac}}g_{ac,b}=0 \ .
\end{align}
Due to Euler-Lagrangian equations (\ref{E-L}) and to the
definitions (\ref{e-m-can}) and (\ref{seym}) of both the
energy-momentum tensors, above formula reduces to the following
statement:
\begin{equation}\label{expr2}
\partial_a {T^a}_b + \frac12
{\tau}^{ac} g_{ac,b} =0\ .
\end{equation}
Our proof of this formula is valid in any coordinate system. In
particular, we may use such a system, for which all partial
derivatives of the metric vanish at a given point $x \in S$. In
this particular coordinate system we have:
\[
{\overline{\nabla} }_a {T^a}_b (x) =\partial_a {T^a}_b (x) =0 \ .
\]
But ${\overline{\nabla} }_a {T^a}_b (x)=0$ is a
coordinate-independent statement: once proved in one coordinate
system, it remains valid in any other system. Repeating this for
all points $x \in S$ separately, we prove Noether theorem
(\ref{Noether}). Subtracting now (\ref{expr2}) from
(\ref{Noether}) we obtain  the following identity:
\[
T^{ab} g_{ab,c} = - {\tau}^{ab} g_{ab,c} \ ,
\]
which must be true in any coordinate system. Here, both $T^{ab}$
and ${\tau}^{ab}$ are defined only up to an additive term of the
form $C X^a X^b$, which vanishes when multiplied by $g_{ab,c}$. In
the standard Riemannian or Lorentzian geometry of a non-degenerate
metric, the derivatives $g_{ab,c}$ may be freely chosen at each
point separately, which immediately implies the
Belinfante-Rosenfeld identity $T = - \tau$. In our case, the
freedom in the choice of these derivatives is restricted by the
constraint. This is the only restriction. Hence, the
Belinfante-Rosenfeld identity is true only up to the annihilator
of these constraints, i.e.,~only in the form of equation
(\ref{R-B}).

\end{proof}

{\bf Remark:} In non-degenerate geometry, vanishing of derivatives
of the metric tensor at a point $x$ uniquely defines a local
``inertial system'' at $x$: if two coordinate systems, say $(x^a)$
and $(y^a)$, fulfill this condition at $x$, then second
derivatives of $x^a$ with respect to $y^b$ {\em vanish}
identically at this point. Covariant derivative may thus be
defined as a partial derivative, but calculated with respect to an
inertial system, i.e.,~to any coordinate system of this class. In
our degenerate case, vanishing of derivatives of the metric does
not fix uniquely the inertial system. There are different
coordinate systems $(x^a)$ and $(y^a)$, for which $g_{ab,c}$
vanishes at $x$, but we have:
\[
\frac {\partial^2 y^a}{\partial x^b x^c } (x) \ne 0 \ .
\]
This is why any attempt to define covariant derivative for an
arbitrary tensor on $S$ fails. This ambiguity is, however,
cancelled by algebraic properties of our energy-momentum tensor,
namely by identities (\ref{G-1}) and (\ref{G-2}). This enables us
to define unambiguously the covariant divergence of
``energy-momentum-like'' tensor-densities using formula
(\ref{div-final}).

\section{Dynamics of the total system ``gravity + shell'':
Lagrangian  version}\label{lagrangian-section}

In this paper we consider dynamics of a light-like matter-shell
discussed in the previous Section, interacting with gravitational
field. We present here a method of derivation of the dynamical
equations of the system, which applies also to a massive shell and
follows the ideas of \cite{shell1}.

The dynamics of the ``gravity + shell'' system will be derived
from the action principle $\delta {\cal A} = 0$, where
\begin{equation}\label{action}
  {\cal A} = {\cal A}_{\mbox{\tiny\rm grav}}^{\mbox{\tiny\rm reg}} +
  {\cal A}_{\mbox{\tiny\rm grav}}^{\mbox{\tiny\rm sing}} +
  {\cal A}_{\mbox{\tiny\rm matter}} \ ,
\end{equation}
is the sum of the gravitational action and the matter action.
Gravitational action, defined as integral of the Hilbert
Lagrangian, splits into the regular and the singular part,
according to decomposition of the curvature:
\begin{align}
  L_{\mbox{\tiny\rm grav}} =   \frac 1{16 \pi} \sqrt{|g|} \ R &=
  \frac 1{16 \pi} \sqrt{|g|} \left(\mbox{\rm reg}(R) +
  \mbox{\rm sing}(R) \right)\nonumber\\& =
   L_{\mbox{\tiny\rm grav}}^{\mbox{\tiny\rm reg}} +
   L_{\mbox{\tiny\rm grav}}^{\mbox{\tiny\rm sing}}
   \ .
\label{Hilbert}
\end{align}
Using formulae (\ref{E-sing-non-deg})--(\ref{Einst-Q}), we express
the singular part of $R$ in terms of the singular part of the
Einstein tensor:
\begin{equation}\label{grav-sing}
  \sqrt{|g|} \
  \mbox{\rm sing}(R) = - \mbox{\rm sing}({\cal G}) =
  - {\bf G}^{\mu\nu} g_{\mu\nu} \boldsymbol\delta(x^3) \ .
\end{equation}
As analyzed in Section \ref{Bianchi}, an additive,
coordinate-dependent ambiguity $C X^\mu X^\nu$ in the definition
of ${\bf G}^{\mu\nu}$ is irrelevant, because cancelled when
contracted with $g_{\mu\nu}$:
\[
{\bf G}^{\mu\nu} g_{\mu\nu} = {\bf G}^{ab} g_{ab}= {{\bf G}^a}_a \
.
\]
For the matter Lagrangian $L_{\mbox{\tiny\rm matter}}$, we assume
that it has properties discussed in the previous Section. Finally,
the total action is the sum of three integrals:
\begin{equation}\label{dod}
    {\cal A} = \int_{D}
    L_{\mbox{\tiny\rm grav}}^{\mbox{\tiny\rm reg}} \
    + \int_{D}
    L_{\mbox{\tiny\rm grav}}^{\mbox{\tiny\rm sing}}  +
    \int_{D \cap S} L_{\mbox{\tiny\rm matter}} \ ,
\end{equation}
where $D$ is a four-dimensional region with boundary in spacetime
$M$ which is possibly cut by a light-like three-dimensional
surface $S$ (actually, because of the Dirac-delta factor, the
second term reduce to integration over ${D \cap S}$). Variation is
taken with respect to the spacetime metric tensor $g_{\mu\nu}$ and
to the matter fields $z^K$ living on $S$. The light-like character
of the matter considered here, implies the light-like character of
$S$ (i.e.,~degeneracy of the induced metric: $\det g_{ab} = 0$) as
an additional constraint imposed on $g$.

We begin with varying the regular part $L_{\mbox{\tiny\rm
grav}}^{\mbox{\tiny\rm reg}}$ of the gravitational action. There
are many ways to calculate variation of the Hilbert Lagrangian.
Here, we use a method proposed by one of us (see \cite{pieszy}).
It is based on the following, simple observation:
\begin{align}
\delta& \left(  \frac 1{16 \pi} \sqrt{|g|} \ g^{\mu\nu} \
R_{\mu\nu} \right)\nonumber\\ & =
 - \frac 1{16 \pi} {\cal G}^{\mu\nu} \delta g_{\mu\nu} +
\frac 1{16 \pi} \sqrt{|g|} \ g^{\mu\nu} \delta R_{\mu\nu}
\label{deltaR}
\end{align}
where
\begin{equation}
{\cal G}^{\mu\nu} := \sqrt{|g|} \ (R^{\mu\nu} - \frac 12
g^{\mu\nu} R) \ .
\end{equation}
It is a matter of a simple algebra, that the last term of
(\ref{deltaR}) is a complete divergence. Namely, the following
formula may be checked by inspection:
\begin{equation} \label{pidR}
{\pi}^{\mu\nu} \delta R_{\mu\nu} =
\partial_\kappa \left( {\pi}_{\lambda}^{\ \mu\nu\kappa} \delta
{\Gamma}^{\lambda}_{\mu\nu} \right)  \ ,
\end{equation}
where we denote
\begin{equation}
{\pi}^{\mu\nu} := \frac 1{16 \pi} \sqrt{|g|} \  g^{\mu\nu} \ ,
\end{equation}
\begin{equation}\label{pi-4}
{\pi}_{\lambda}^{\ \mu\nu\kappa} := {\pi}^{\mu\nu}
\delta^\kappa_\lambda - {\pi}^{\kappa ( \nu} \delta^{\mu
)}_\lambda \ ,
\end{equation}
and ${\Gamma}^{\lambda}_{\mu\nu}$ {\em are not} independent
quantities, but the Christoffel symbols, i.e.,~combinations of the
metric components $g_{\mu\nu}$ and their derivatives. In the above
calculations we use that fact that the covariant derivative
$\nabla {\pi}$ of $\pi$ with respect to $\Gamma$ vanishes
identically, i.e.,~that the following identity holds:
\begin{equation}\label{delta-pi-4}
  \partial_\kappa {\pi}_{\lambda}^{\ \mu\nu\kappa} \equiv
  {\pi}_{\alpha}^{\ \mu\nu\kappa} {\Gamma}^{\alpha}_{\lambda\kappa}
  -{\pi}_{\lambda}^{\ \alpha\nu\kappa} {\Gamma}^{\mu}_{\alpha\kappa}
  -{\pi}_{\lambda}^{\ \mu\alpha\kappa} {\Gamma}^{\nu}_{\alpha\kappa}
  \ .
\end{equation}
Hence, for the regular part of the curvature we obtain:
\begin{equation}\label{action-g}
  \delta \left(  \frac 1{16 \pi} \sqrt{|g|} \
   R \right) =
   - \frac 1{16 \pi} {\cal G}^{\mu\nu} \delta g_{\mu\nu} +
   \partial_\kappa \left( {\pi}_{\lambda}^{\ \mu\nu\kappa} \delta
   {\Gamma}^{\lambda}_{\mu\nu} \right)  \ .
\end{equation}
We shall integrate the above equation over both parts $D^+$ and
$D^-$ of $D$, resulting from cutting $D$ with the surface $S$.
This way we obtain:
\begin{equation}\label{delta-A-grav}
  \delta L_{\mbox{\tiny\rm grav}}^{\mbox{\tiny\rm reg}} =
   - \frac 1{16 \pi}
   \mbox{\rm reg}({\cal G})^{\mu\nu} \delta g_{\mu\nu} +
   \mbox{\rm reg}\left(
   \partial_\kappa \left( {\pi}_{\lambda}^{\ \mu\nu\kappa} \delta
   {\Gamma}^{\lambda}_{\mu\nu} \right) \right)  \ .
\end{equation}
Now, we are going to prove that the analogous formula is valid
also for the singular part of the gravitational Lagrangian,
i.e.,~that the following formula holds:
\begin{equation}\label{L-grav-sing20}
  \delta L_{\mbox{\tiny\rm grav}}^{\mbox{\tiny\rm sing}} =
  - \frac 1{16 \pi} \mbox{\rm sing}({\cal G})^{\mu\nu}
   \delta  g_{\mu\nu}  +
   \mbox{\rm sing}\left(
   \partial_\kappa \left( {\pi}_{\lambda}^{\ \mu\nu\kappa} \delta
   {\Gamma}^{\lambda}_{\mu\nu} \right) \right)
   \ .
\end{equation}
To prove this formula, we calculate the singular part of the
divergence $\partial_\kappa \left( {\pi}_{\lambda}^{\
\mu\nu\kappa} \delta {\Gamma}^{\lambda}_{\mu\nu} \right)$. Because
all these quantities are invariant, geometric objects
($\delta\Gamma$ is a tensor!), we may calculate them in an
arbitrary coordinate system. Hence, we may use our adapted
coordinate system described in previous Sections, where coordinate
$x^3$ is constant on $S$. This way, using (\ref{pi-4}), we obtain:
\begin{align}\label{delta-S}
\mbox{\rm sing}\left(
   \partial_\kappa \left( {\pi}_{\lambda}^{\ \mu\nu\kappa} \delta
   {\Gamma}^{\lambda}_{\mu\nu} \right) \right) &=
  {\pi}_{\lambda}^{\ \mu\nu\perp} \delta
   [ {\Gamma}^{\lambda}_{\mu\nu} ] =
   {\pi}_{\lambda}^{\ \mu\nu 3} \delta
   [ {\Gamma}^{\lambda}_{\mu\nu} ]\nonumber\\& =
   {\pi}^{\mu\nu} \delta
   [ A^{3}_{\mu\nu} ] \ ,
\end{align}
where by $A$ we denote:
\begin{equation} \label{AG}
A^{\lambda}_{\mu\nu} := {\Gamma}^{\lambda}_{\mu\nu} -
{\delta}^{\lambda}_{(\mu} {\Gamma}^{\kappa}_{\nu ) \kappa}
\end{equation}
(Do not try to attribute any sophisticated geometric
interpretation to $A^{\lambda}_{\mu\nu}$; it is merely a
combination of the Christoffel symbols, which arises frequently in
our calculations. It has been introduced for technical reasons
only.) The following combination of the connection coefficients
will also be useful in the sequel:
\begin{equation} \label{tQ}
{\widetilde  Q}^{\mu\nu} := \sqrt{|g|} \left( g^{\mu \alpha}
g^{\nu \beta} - \frac 12 g^{\mu\nu} g^{ \alpha \beta} \right)
A^3_{\alpha\beta}  \ .
\end{equation}
It may be immediately checked that:
\begin{equation}\label{A-przez-Q}
  {\pi}^{\mu\nu} \delta A^{3}_{\mu\nu} = - \frac 1{16 \pi}
  g_{\mu\nu} \delta {\widetilde  Q}^{\mu\nu}  \ .
\end{equation}
In Appendix \ref{GuuA} we analyze in detail the structure of
quantity ${\widetilde Q}$. As a combination of the connection
coefficients, it {\em does not} define any tensor density. But it
differs from the external curvature $Q(X)$ of $S$ introduced in
Section \ref{geometry2}, only by terms containing metric
components and their derivatives {\em along} $S$. Jumps of these
terms across $S$ vanish identically. Hence, the following is true:
\begin{equation}\label{tilde-Q}
  [{\widetilde  Q}^{\mu\nu}]
  \boldsymbol\delta(x^3)
   = [Q^{\mu\nu}]
  \boldsymbol\delta(x^3) =  \mbox{\rm sing}({\cal G})^{\mu\nu}
  \ .
\end{equation}
Consequently, formulae (\ref{delta-S}), (\ref{A-przez-Q}) and
(\ref{grav-sing}) imply:
\begin{align}\label{Lgrav-sing}
    \boldsymbol\delta(x^3)
   {\pi}_{\lambda}^{\ \mu\nu\perp} \delta
   [ {\Gamma}^{\lambda}_{\mu\nu} ]& = -
   \frac 1{16 \pi} g_{\mu\nu} \delta \
   \mbox{\rm sing}({\cal G})^{\mu\nu} \nonumber\\ & =
    L_{\mbox{\tiny\rm grav}}^{\mbox{\tiny\rm sing}}
   +\frac 1{16 \pi} \mbox{\rm sing}({\cal G})^{\mu\nu}
   \delta  g_{\mu\nu}
   \ ,
\end{align}
which ends the proof of (\ref{L-grav-sing20}). Summing up
(\ref{delta-A-grav}) and (\ref{L-grav-sing20}) we obtain:
\begin{equation}\label{delta-A-grav-k}
  \delta L_{\mbox{\tiny\rm grav}} =
  - \frac 1{16 \pi}
   {\cal G}^{\mu\nu} \delta g_{\mu\nu} +
   \partial_\kappa \left( {\pi}_{\lambda}^{\ \mu\nu\kappa} \delta
   {\Gamma}^{\lambda}_{\mu\nu} \right)\ ,
\end{equation}
where both terms are composed of its regular and singular
part\footnote{In \cite{pieszy} the above formula was proved for
regular spacetimes. In \cite{shell1} its validity was extended to
spacetimes with a three-dimensional, non-degenerate
curvature-singularity. Here, we have shown that it is also valid
for a light-like curvature-singularity.}.

Now, we calculate the variation of the matter part
$L_{\mbox{\tiny\rm matter}}$  of the action on $S$:
\begin{eqnarray}
  \delta L_{\mbox{\tiny\rm matter}} & = & \frac
  {\partial L_{\mbox{\tiny\rm mater}}}{\partial g_{ab}} \delta
  g_{ab} + \frac {\partial L_{\mbox{\tiny\rm matter}}}{\partial z^K}
  \delta z^K + \frac
  {\partial L_{\mbox{\tiny\rm matter}}}{\partial {z^K}_a}
  \partial_a \delta z^K \nonumber \\
  & = &  \frac 12 \tau^{ab} \delta g_{ab}
  + \left( \frac {\partial
  L_{\mbox{\tiny\rm matter}}}{\partial z^K} - \partial_a
  \frac
  {\partial L_{\mbox{\tiny\rm matter}}}{\partial {z^K}_a}
  \right) \delta z^K  \nonumber \\ &  & +
  \partial_a \left( {p_K}^a \delta z^K \right) \label{delta-L-matter}
  \ ,
\end{eqnarray}
where we used definition (\ref{seym}) and have introduced the
momentum canonically conjugate to the matter variable $z^K$:
\begin{equation}\label{momentum-p}
  {p_K}^a := \frac
  {\partial L_{\mbox{\tiny\rm matter}}}{\partial {z^K}_a} \ .
\end{equation}
Finally, we obtain the following formula for the variation of the
total (``matter + gravity'') Lagrangian:
\begin{align}\label{Lag-tot-var}
  \delta L  = &  - \frac 1{16 \pi}
   \mbox{\rm reg}({\cal G})^{\mu\nu} \delta g_{\mu\nu} \nonumber\\&+
   \boldsymbol\delta(x^3)
   \left( \frac {\partial
  L_{\mbox{\tiny\rm matter}}}{\partial z^K} - \partial_a
  \frac
  {\partial L_{\mbox{\tiny\rm matter}}}{\partial {z^K}_a}
  \right) \delta z^K \nonumber \\
  & -  \boldsymbol\delta(x^3)
   \frac 1{16 \pi} \left( {\bf G}^{ab} - 8 \pi
     \tau^{ab}  \right) \delta g_{ab} \nonumber\\
  & +
   \partial_\kappa \left( {\pi}_{\lambda}^{\ \mu\nu\kappa} \delta
   {\Gamma}^{\lambda}_{\mu\nu} \right) +
   \boldsymbol\delta(x^3)
   \partial_a \left( {p_K}^a \delta z^K \right)
   \ .
\end{align}
In this Section we assume that both $\delta g_{\mu\nu}$ and
$\delta z^K$ vanish in a neighborhood of the boundary $\partial D$
of the spacetime region $D$ (this assumption will be later
relaxed, when deriving Hamiltonian structure of the theory).
Hence, the last two boundary terms of the above formula vanish
when integrated over $D$. Vanishing of the variation $\delta {\cal
A} = 0$ with fixed boundary values implies, therefore, the
Euler-Lagrange equations (\ref{E-L}) for the matter field $z^K$,
together with Einstein equations for gravitational field. Regular
part of Einstein equations:
\[
\mbox{\rm reg}({\cal G})^{\mu\nu} = 0
\]
must be satisfied outside of $S$ and the singular part must be
fulfilled on $S$. To avoid irrelevant ambiguities of the type $C
X^a X^b$, we write it in the following form, equivalent to the
Barrab\`es-Israel equation:
\begin{equation}
{{\bf G}^a}_b = 8 \pi {\tau^a}_b \ .
\end{equation}
Summing up singular and regular parts of the above quantities we
may write the ``total Einstein equations'' in the following way:
\begin{align}\label{Lag-111}
  \delta L  = &  \frac 1{16 \pi}
   \left({\cal G}^{\mu\nu}  - 8 \pi
     {\cal T}^{\mu\nu}  \right)\delta g_{\mu\nu} \nonumber\\&+
   \boldsymbol\delta(x^3)
   \left( \frac {\partial
  L_{\mbox{\tiny\rm matter}}}{\partial z^K} - \partial_a
  \frac
  {\partial L_{\mbox{\tiny\rm matter}}}{\partial {z^K}_a}
  \right) \delta z^K \nonumber\\
  & +
   \partial_\kappa \left( {\pi}_{\lambda}^{\ \mu\nu\kappa} \delta
   {\Gamma}^{\lambda}_{\mu\nu} \right) +
   \boldsymbol\delta(x^3)
   \partial_a \left( {p_K}^a \delta z^K \right)
   \ .
\end{align}
Here, we have defined the four-dimensional energy-momentum tensor:
${\cal T}^{\mu\nu} := \boldsymbol\delta(x^3) \tau^{\mu\nu}$ with
$\tau^{3\nu} \equiv 0$. Since $\tau^{ab}$ was defined up to an
additive term $C X^a X^b$, this ambiguity remains and ${\cal
T}^{\mu\nu}$ is defined up to $C X^\mu X^\nu
\boldsymbol\delta(x^3)$, similarly as the quantity ${\cal
G}^{\mu\nu}$. This ambiguity is annihilated when contracted with
$\delta g_{\mu\nu}$.

\section{Dynamics of the total system ``gravity + shell'':
Hamiltonian description}\label{hamiltonian-section}

Field equations of the theory (Euler-Lagrange equations for matter
and Einstein equations---both singular and regular---for gravity)
may thus be written in the following way\footnote{Formula
(\ref{divergence}) is analogous to formula: $dL(q,\dot{q}) =
\left( p dq \right)\dot{} = \dot{p} dq + p d \dot{q}$ in
mechanics, which contains both the dynamical equation: $\dot{p} =
\partial L / \partial q$, and the definition of the canonical
momentum: $p = \partial L / \partial \dot{q}$. For detailed
analysis of this structure see \cite{pieszy}.}:
\begin{equation}\label{divergence}
  \delta L =
   \partial_\kappa \left( {\pi}_{\lambda}^{\ \mu\nu\kappa} \delta
   {\Gamma}^{\lambda}_{\mu\nu} \right) +
   \boldsymbol\delta(x^3)
   \partial_a \left( {p_K}^a \delta z^K \right)
   \ .
\end{equation}
Indeed, field equations are equivalent to the fact that the volume
terms (\ref{Lag-111}) in the variation of the Lagrangian must
vanish identically. Hence, the entire dynamics of the theory of
the system ``matter + gravity'' is equivalent to the demand, that
variation of the Lagrangian is equal to boundary terms only.
Similarly as in equation (\ref{delta-S}), we may use definition of
${\pi}_{\lambda}^{\ \mu\nu\kappa}$ and express it in terms of the
contravariant density of metric $\pi^{\mu\nu}$. This way we
obtain:
\begin{equation}
   {\pi}_{\lambda}^{\ \mu\nu \kappa} \delta
   {\Gamma}^{\lambda}_{\mu\nu}=\pi^{\mu\nu}\delta
   A^\kappa_{\mu\nu}\ .
\end{equation}
Hence, field equations may be written in the following way:
\begin{equation}\label{divergence-A}
  \delta L =
   \partial_\kappa \left( \pi^{\mu\nu}\delta
   A^\kappa_{\mu\nu} \right) +
   \boldsymbol\delta(x^3)
   \partial_a \left( {p_K}^a \delta z^K \right)
   \ .
\end{equation}
As soon as we choose a (3+1)-decomposition of the spacetime $M$,
our field theory will be converted into a Hamiltonian system, with
the space of Cauchy data on each of the three-dimensional surfaces
playing role of an infinite-dimensional phase space. Let us choose
coordinate system adapted to this (3+1)-decomposition. This means
that the time variable $t = x^0$ is constant on three-dimensional
surfaces of this foliation. We assume that these surfaces are
space-like. To obtain Hamiltonian formulation of our theory we
shall simply integrate equation (\ref{divergence})
(or---equivalently---(\ref{divergence-A})) over such a Cauchy
surface $\Sigma_t \subset M$ and then perform Legendre
transformation between time derivatives and corresponding momenta.

In the present paper we consider the case of an asymptotically
flat spacetime and assume that also leaves $\Sigma_t$ of our
(3+1)-decomposition are asymptotically flat at infinity. To keep
control over 2-dimensional surface integrals at spatial infinity,
we first consider dynamics of our ``matter + gravity'' system in a
finite world tube ${\cal U}$, whose boundary carries a
non-degenerate metric of signature $(-,+,+)$. At the end of our
calculations, we shift the boundary $\partial {\cal U}$ of the
tube to space-infinity. We assume that the tube contains the
surface $S$ together with our light-like matter travelling over
it.

Denoting by $V := {\cal U} \cap \Sigma_t$ the portion of
$\Sigma_t$ which is contained in the tube ${\cal U}$, we thus
integrate (\ref{divergence-A}) over the finite volume $V \subset
\Sigma_t$ and keep surface integrals on the boundary $\partial V$
of $V$. They will produce the ADM~mass as the Hamiltonian of the
total ``matter + gravity'' system at the end of our calculations,
when we pass to infinity with $\partial V = \Sigma_t \cap \partial
{\cal U}$. Because our approach is geometric and does not depend
upon the choice of coordinate system, we may further simplify our
calculations using coordinate $x^3$ adapted to both $S$ and to the
boundary $\partial {\cal U}$ of the tube. We thus assume that
$x^3$ is constant on both these surfaces.

Integrating (\ref{divergence-A}) over the volume $V$ we thus
obtain:
\begin{align}
\label{generate-20}
  \delta&\int_V  L =\int_V \partial_\kappa \left(
  \pi^{\mu\nu}\delta
   A^\kappa_{\mu\nu} \right) +
   \int_V \boldsymbol\delta(x^3)
   \partial_a \left( {p_K}^a \delta z^K \right)\nonumber \\
  & =\int_V\left( \pi^{\mu\nu}\delta
   A^0_{\mu\nu} \right)^\cdot + \int_{\partial V}
   \pi^{\mu\nu}\delta
   A^\perp_{\mu\nu}
+\int_{V\cap S}
    \left( {p_K}^0 \delta z^K \right)^\cdot \ ,
\end{align}
where by ``dot'' we denote time derivative. In the above formula
we have skipped the two-dimensional divergencies which vanish when
integrated over surfaces $\partial V$ and $V\cap S$.

To further simplify our formalism, we denote by $p_K := {p_K}^0$
the time-like component of the momentum canonically conjugate to
the field variable $z^K$ and perform the Legendre transformation:
\begin{equation}\label{legender-matter}
  \left( p_K \delta z^K \right)^\cdot = \dot{p}_K \delta z^K -
  \dot{z}^K \delta p_K + \delta \left( p_K \dot{z}^K \right) \ .
\end{equation}
The last term, put on the left-hand side of (\ref{generate-20}),
meets the matter Lagrangian and produces the matter Hamiltonian
(with minus sign), according to formula:
\begin{equation}\label{T00}
  L_{\mbox{\tiny\rm matter}} - p_K \dot{z}^K =
  L_{\mbox{\tiny\rm matter}} - {p_K}^0 {z^K}_0 = - {T^0}_0 =
  {\tau^0}_0 \ .
\end{equation}
To perform also Legendre transformation in gravitational degrees
of freedom we follow here method proposed by one of us (see
\cite{pieszy}). For this purpose we first observe that, due to
metricity of the connection $\Gamma$, the gravitational
counterpart $\pi^{\mu\nu}\delta A^0_{\mu\nu}$ of the canonical
one-form $p_K \delta z^K$ reduces as follows:
\begin{equation}\label{redukcja-0}
    \pi^{\mu\nu}\delta A^0_{\mu\nu}=
    -\frac{1}{16\pi}g_{kl}\delta P^{kl} +
    \partial_k\left(\pi^{00}\delta\left(
    \frac{\pi^{0k}}{\pi^{00}}\right)\right) \ ,
\end{equation}
where $P^{kl}$ denotes the external curvature of $\Sigma$ written
in the ADM~form. Similarly, the boundary term $\pi^{\mu\nu}\delta
A^\perp_{\mu\nu}=\pi^{\mu\nu}\delta A^3_{\mu\nu}$ reduces as
follows:
\begin{equation}\label{redukcja-3}
    \pi^{\mu\nu}\delta A^3_{\mu\nu}=
    -\frac{1}{16\pi}g_{ab}\delta {\cal Q}^{ab} +
    \partial_a\left(\pi^{33}\delta\left(
    \frac{\pi^{3a}}{\pi^{33}}\right)\right) \ ,
\end{equation}
where ${\cal Q}^{ab}$ denotes the external curvature of the tube
$\partial {\cal U}$ written in the ADM~form. A simple proof of
these formulae is given in Appendix \ref{ap-redukcja}.

Using these results and skipping the two-dimensional divergencies
which vanish after integration, we may rewrite gravitational part
of (\ref{generate-20}) in the following way:
\begin{align}\label{generate-grav}
   \int_V&\left( \pi^{\mu\nu}\delta
   A^0_{\mu\nu} \right)^\cdot + \int_{\partial V}
   \pi^{\mu\nu}\delta
   A^\perp_{\mu\nu} \nonumber \\  = &
   -\frac{1}{16\pi}\int_V\left( g_{kl}\delta P^{kl} \right)^\cdot
   -\frac{1}{16\pi} \int_{\partial V} g_{ab}\delta {\cal Q}^{ab}\nonumber\\
   & +  \int_{\partial V} \left(\pi^{00}\delta\left(
    \frac{\pi^{03}}{\pi^{00}}\right) + \pi^{33}\delta\left(
    \frac{\pi^{30}}{\pi^{33}}\right) \right)^\cdot \ .
\end{align}
The last integral may be rewritten in terms of the hyperbolic
angle $\alpha$ between surfaces $\Sigma$ and $\partial {\cal U}$,
defined as: $\alpha  = {\rm arcsinh} (q)$, where
\begin{equation}
q =\frac{g^{30}}{\sqrt{| g^{00}g^{33}|}}\ ,
\end{equation}
and the two-dimensional volume form $\lambda = \sqrt{\det g_{AB}}$
on $\partial V$, in the following way:
\begin{equation}\label{lambda-alpha}
  \pi^{00}\delta\left(
    \frac{\pi^{03}}{\pi^{00}}\right) + \pi^{33}\delta\left(
    \frac{\pi^{30}}{\pi^{33}}\right) = \frac{1}{8\pi} \lambda
    \delta \alpha \ .
\end{equation}
For the proof of this formula see Appendix \ref{ap-lambda-alpha}.
Hence, we have:
\begin{align}
 \int_{V}&\left(  \pi^{\mu\nu}\delta A_{\mu\nu}^{0}\right)  ^{\cdot}%
+\int_{\partial V}\pi^{\mu\nu}\delta A_{\mu\nu}^{\perp} \label{generate-grav1}%
\\
=&-\frac{1}{16\pi}\int_{V}\left(  g_{kl}\delta P^{kl}\right)  ^{\cdot}%
-\frac{1}{16\pi}\int_{\partial
V}g_{ab}\delta\mathcal{Q}^{ab}\nonumber\\&+\frac{1}{8\pi
}\int_{\partial V}\left(  \lambda\delta\alpha\right)  ^{\cdot}
\nonumber\ .
\end{align}
Now we perform the Legendre transformation both in the volume:
\[
\left(  g_{kl}\delta P^{kl}\right)  ^{\cdot}=\left(
\dot{g}_{kl}\delta
P^{kl}-\dot{P}^{kl}\delta g_{kl}\right)  +\delta\left(  g_{kl}\dot{P}%
^{kl}\right)
\]
and on the boundary:
$(\lambda\delta\alpha)^{\cdot}=(\dot{\lambda}\delta
\alpha-\dot{\alpha}\delta\lambda)+\delta(\lambda\dot{\alpha})$.

In appendix \ref{ap-pe-lambda} we prove  the following formula:
\begin{align}\label{pe-lambda1}
-\frac{1}{16\pi}& \int_{V}\left(  g_{kl}\dot P^{kl}\right) +\frac
{1}{8\pi} \int_{\partial V}\lambda\dot{\alpha}\nonumber\\
=&\frac{1}{8\pi}  \int
_{V}\sqrt{|g|}R^0{_0}+\frac{1}{16\pi}\int_{\partial V} \left(
\mathcal{Q}^{AB}g_{AB}-\mathcal{Q}^{00}g_{00}\right)\ ,
\end{align}
Then, we have
\begin{align}
\frac{1}{16\pi} &\int_V L_{\mbox{\tiny\rm grav}} - \frac{1}{8\pi}
\int_V \sqrt{|g|}R^0{_0}\nonumber\\& =\frac{1}{8\pi} \int_V
\sqrt{|g|}\left(\frac 12 R - R^0{_0}\right)= -\frac{1}{8\pi}
\int_V {\cal G}^0{_0}\ .
\end{align}
Splitting the component ${\cal G}^0{_0}$ of the Einstein tensor
into regular and singular part we obtain
\begin{equation}
\frac{1}{8\pi} \int_V {\cal G}^0{_0} =\frac{1}{8\pi} \int_V{\rm
reg}\left( {\cal G}^0{_0}\right) +\frac{1}{8\pi} \int_V{\rm
sing}\left( {\cal G}^0{_0}\right) \ ,
\end{equation}
The regular part of Einstein tensor density ${\rm reg}\left({\cal
G}^{\mu\nu}\right)$ vanishes due to field equations. The singular
part:
\begin{equation}
{\rm sing}\left( {\cal G}^0{_0}\right)={\boldsymbol
\delta}(x^3){\bf G}^0{_0} \ ,
\end{equation}
meets the matter hamiltonian $\tau^0{_0}$ (see formula
(\ref{T00})) and gets annihilated due to Einstein equations:
\begin{equation}
 \frac{1}{8\pi} \int_{V
\cap S}\left( {\bf G}^0{_0}-8\pi\tau^0{_0}\right)=0\ .
\end{equation}
Finally, we obtain the following generating formula (cf.
\cite{pieszy}):
\begin{align}
0   =&\frac{1}{16\pi}\int_{V}\left(  \dot{P}^{kl}\delta g_{kl}-\dot{g}%
_{kl}\delta P^{kl}\right)  +\frac{1}{16\pi}\int_{\partial V}(
\dot{\lambda}\delta\alpha
  -\dot{\alpha}\delta\lambda)
 \nonumber \\
&  +\int_{V\cap S}\left(  \dot{p}^{0}{_{K}}\delta
z^{K}-\dot{z}^{K}\delta p^{0}{_{K}}\right)
-\frac{1}{16\pi}\int_{\partial V}%
g_{ab}\delta\mathcal{Q}^{ab}
\nonumber\\&+\frac{1}{16\pi}\delta\int_{\partial V}\left(
\mathcal{Q}^{AB}g_{AB}-\mathcal{Q}^{00}g_{00}\right)
 \ .
\end{align}

Using results of \cite{pieszy} it may be easily shown that pushing
the boundary $\partial V$ to infinity and handling in a proper way
the above three surface integrals over $\partial V$, one obtains
in the asymptotically flat case the standard Hamiltonian formula
for both gravitational and matter degrees of freedom, with the
ADM~mass (given by the resulting surface integral at infinity)
playing role of the total Hamiltonian. More precisely, denoting
the matter momenta by
\begin{equation}\label{pi-matter}
  \pi_K := p^{0}{_{K}} {\boldsymbol \delta}(x^3) \ ,
\end{equation}
the final formula for $\partial V \rightarrow \infty$ reads:
\begin{align}
- \delta {\cal H}  =  \frac 1{16 \pi} \int_V & \left( {\dot
P}^{kl} \delta g_{kl} - {\dot g}_{kl} \delta P^{kl}
\right)\nonumber\\ &\qquad+\int_{V}\left(  \dot{\pi}_K\delta
z^{K}-\dot{z}^{K}\delta \pi_K\right) \label{dbarH-grav} \ ,
\end{align}
where ${\cal H}$ is the ``total hamiltonian'', equal to the
ADM~mass at spatial infinity\footnote{Formula (\ref{dbarH-grav})
is analogous to formula: $-dH(q,p) = \dot{p} dq -  \dot{q} dp$ in
mechanics. In a non-constrained case this formula is equivalent to
the definition of the Hamiltonian vector field $(\dot{p},\dot{q})$
{\em via} Hamilton equations: $\dot{p} =- \partial H / \partial
q$, and $\dot{q} =
\partial H / \partial p $. We stress, however, that the formula is
much more general and is valid also for constrained systems, when
the field is not unique, but given only ``up to a gauge''. For
detailed analysis of this structure see \cite{pieszy}.}.

\section{Constraints}\label{constraints-section}

Consider Cauchy data $(P^{kl} , g_{kl} , \pi_K , z^K)$ on a
three-dimensional space-like surface $V_{t}$ and denote by
$\gt^{kl}$ the three-dimensional metric inverse to $g_{kl}$.
Moreover, we use the following notation: $\sgth:=\sqrt{\det
g_{kl}}$, $\stackrel{(3)}{R}$ is the three-dimensional scalar
curvature of $g_{kl}$, $P:=P^{kl}  g_{kl}$ and ``$|$'' is the
three-dimensional covariant derivative with respect to $g_{kl}$.

We are going to prove that these data must fulfill constraints
implied by Gauss-Codazzi equations for the components ${\cal
G}^0{_\mu}$ of the Einstein tensor density. Standard decomposition
of ${\cal G}^0{_\mu}$ into the spatial (tangent to $V_{t}$) part
and the time-like (normal to $V_{t}$) part gives us respectively:
\begin{equation}\label{GCww}
{\cal G}^0{_l} =- P_l{^k}{_{|k}}\ ,
\end{equation}
and
\begin{equation}\label{GCws}
 2{\cal G}^0{_\mu} n^{\mu} =
-
\sgth\stackrel{(3)}{R}+\left(P^{kl}P_{kl}-\frac{1}{2}
P^2\right)\sgt  \, .
\end{equation}
Here by $n$ we have denoted the future orthonormal vector to
Cauchy surface $V_t$:
\[
n^\mu=- \frac {g^{0\mu}}{\sqrt{-g^{00}}}\ .
\]
Vacuum Einstein equations outside and inside of $S$ imply
vanishing of the regular part of ${\cal G}^0{_\mu}$. Hence, the
regular part of the vector constraint reads:
\[ {\rm reg}\left( P_l{^k}{_{|k}}\right)=0 \ , \]
whereas the regular part of the scalar constraint reduces to:
\[  {\rm reg} \left(
\sgth\stackrel{(3)}{R} - \left(P^{kl}P_{kl}-\frac{1}{2}
P^2\right)\sgt  \right)= 0 \, .\] The singular part of
constraints, with support on the intersection sphere $S_{t} =
V_{t} \cap S$, can be derived as follows.

Singular part of three dimensional derivatives of the ADM~momentum
$P_{kl}$ consists of derivatives in the direction of $x^3$:
\[
{\rm sing}(P_l{^k}_{|k}) ={\rm sing}(\partial_3 P_l{^3})=
\boldsymbol\delta(x^3)[P_l{^3}]\ ,
\]
so the full vector constraint has the form
\begin{equation}\label{wwd}
  P_l{^k}_{|k} = [P_l{^3}] \boldsymbol\delta(x^3)\ .
\end{equation}
Components of the ADM~momentum  $P^{kl}$ are regular, hence
singular part of the term $\left(P^{kl}P_{kl}-\frac{1}{2}
P^2\right)$ vanishes. Singular part of the three-dimensional
scalar curvature consists of derivatives in the direction of $x^3$
of the (three-dimensional) connection coefficients:
\begin{align}
 {\rm sing}(\stackrel{(3)}{R})&= {\rm sing} \left(
 \partial_3 \bigl( \Gamma^3_{kl}\gt^{kl}
 -\Gamma^m_{ml}\gt^{3l} \bigr)\right)\nonumber\\
& =
 \boldsymbol\delta(x^3)\left[\Gamma^3_{kl}\gt^{kl}
 -\Gamma^m_{ml}\gt^{3l}\right]\ ,
\end{align}
and expression in the square brackets may be reduced to the
following term
\begin{align}
\sgth \left[\Gamma^3_{kl}\gt^{kl} -\Gamma^m_{ml}\gt^{3l}\right]&=
-2\sqrt{\gt^{33}}
\left[\partial_3\biggl(\sgth\sqrt{\gt^{33}}\biggr)\right]\nonumber\\
&= -2\sqrt{\gt^{33}} \left[ \partial_k\left(\frac{\sgth
\gt^{3k}}{\sqrt{\gt^{33}}}\right)\right]  \, ,
\end{align}
because derivatives tangent to  $S$ are continuous. But expression
in square brackets is equal to the external curvature scalar $k$
for the two-dimensional surface $S_t \subset V_t$:
\begin{equation}
\sgth k= -\partial_k\left(\frac{\sgth
\gt^{3k}}{\sqrt{\gt^{33}}}\right) \, .
\end{equation}
So we get
\[
{\rm sing}\left( \sgth \stackrel{(3)}{R}\right)= 2 \sgth
{\sqrt{\gt^{33}}}  [k] \boldsymbol\delta(x^3) = 2 [\lambda k]
\boldsymbol\delta(x^3) \, ,\] and finally:
\begin{equation}\label{wsd}
\sgth\stackrel{(3)}{R} - \left(P^{kl}P_{kl}-\frac{1}{2}
P^2\right)\sgt
  = 2 [\lambda k] \boldsymbol\delta(x^3) \, .
\end{equation}
Equations (\ref{wwd}) and (\ref{wsd}) give a generalization (in
the sense of distributions) of the usual vacuum constraints
(vector and scalar respectively).

Now, we will show how the distributional matter located on $S_t$
determines the four surface quantities $[P^3_{\ k}]$ and $[\lambda
k]$, entering into the singular part of the constraints. The
tangent (to $S$) part of ${\cal G}^0{_\mu}$ splits into the
two-dimensional part tangent to $S_t$ and the transversal part
(along null rays).

The tangent to $S_{t}$ part of Einstein equations gives the
following:
\begin{equation}
{\cal G}^0{_A}=8\pi \boldsymbol\delta (x^3) \tau^0{_A} \ ,
\end{equation}
which, due to (\ref{GCww}) and (\ref{wwd}), implies the following
two constraints:
\begin{equation}\label{PBt}
\left[ P^3{_B}\right]=-8\pi\tau^0{_B}\ .
\end{equation}

The remaining null tangent part of Einstein equations reads:
\begin{equation}\label{GXt}
{\cal G}^0{_\mu}X^\mu  =8\pi \boldsymbol\delta
(x^3)\tau^0{_\mu}X^\mu=0\ ,
\end{equation}
because $\tau^0{_\mu}X^\mu=0$. In Appendix \ref{dPk} we show that
this equation reduces to the following constraint:
\begin{equation} \label{plk}
\left[\frac{P^{33}}{\sqrt{\gt^{33}}}+ \lambda k \right]=0\ .
\end{equation}

We remind the reader that the singular part of ${\cal G}^0{_3}$
cannot be defined in any intrinsic way. Consequently, we have only
three constraints for the singular part (\ref{plk}) and
(\ref{PBt}). The fourth constraint (in a non-degenerate case) has
been replaced here by the degeneracy condition $\det g_{ab}$ for
the metric on $S$. Equations (\ref{PBt}), (\ref{plk}) together
with (\ref{wwd}) and (\ref{wsd}) are the initial value
constraints.

\begin{acknowledgments}
The authors are much indebted to Petr H\'{a}j\'{\i}\v{c}ek and
Jerzy Lewandowski for inspiring discussions. This work was
supported in part by the Polish KBN Grant Nr. 2 P03A 047 15.
\end{acknowledgments}

\appendix
\section{Structure of the singular Einstein tensor}
\label{GuuA} 

 \newcommand{\abar}{\chi}
 \newcommand{\tQ}{{\widetilde Q}}
We rewrite the Ricci tensor:
\begin{equation}\label{aric1}
 R_{\mu\nu}=\partial_\lambda{\Gamma}^{\lambda}_{\mu\nu} - \partial_{(\mu}
{\Gamma}^{\lambda}_{\nu )
\lambda}+{\Gamma}^{\lambda}_{\sigma\lambda}
{\Gamma}^{\sigma}_{\mu\nu} - {\Gamma}^{\lambda}_{\mu\sigma}
{\Gamma}^{\sigma}_{\nu\lambda} \ ,
\end{equation}
in terms of the following combinations of Christoffel symbols (cf.
(\ref{AG}) in Section 5):
\begin{equation}\label{A-def}
A^{\lambda}_{\mu\nu} := {\Gamma}^{\lambda}_{\mu\nu} -
{\delta}^{\lambda}_{(\mu} {\Gamma}^{\kappa}_{\nu ) \kappa} \ .
\end{equation}
We have:
\begin{equation}
R_{\mu\nu}=\partial_\lambda
A^{\lambda}_{\mu\nu}-A^{\lambda}_{\mu\sigma}
A^{\sigma}_{\nu\lambda} + \frac 13 A^{\lambda}_{\mu\lambda}
A^{\sigma}_{\nu\sigma}.
\end{equation}

Terms quadratic in $A$'s may have only step-like discontinuities.
The derivatives along $S$ are thus bounded and belong to the
regular part of the Ricci tensor. The singular part of the Ricci
tensor is obtained from the transversal derivatives only. In our
adapted coordinate system, where $x^3$ is constant on $S$, we
obtain:
\begin{equation}
{\rm sing}(R_{\mu\nu})=\partial_3
A^3_{\mu\nu}=\boldsymbol\delta(x^3)[A^3_{\mu\nu}] \ ,
\end{equation}
where by $\boldsymbol\delta$ we denote the Dirac
delta-distribution and by square brackets we denote the jump of
the value of the corresponding expression between the two sides of
$S$. Consequently, the singular part of Einstein tensor density
reads:
\begin{equation}\label{Einstein}
{\rm sing}({{\cal G}^\mu}_\nu) := \sqrt{|g|} \  {\rm sing} \left(
{ R^\mu}_\nu - \frac 12 R \right)  =  \boldsymbol\delta(x^3) {{\bf
G}^\mu}_\nu\ ,
\end{equation}
where
\begin{equation}\label{G-grube}
{{\bf G}^\mu}_\nu  :=  \sqrt{|g|} \left(\delta^\beta_\nu g^{\mu
\alpha}  - \frac 12  \delta^\mu_\nu g^{ \alpha \beta} \right)
[A^3_{\alpha\beta}] = [\tQ^{\mu}{_\nu}] \ .
\end{equation}
We shall prove that the contravariant version of this quantity:
\[ {\rm sing}({\cal G})^{\mu\nu} =
[\tQ^{\mu \nu}]\boldsymbol\delta(x^3) \ ,
\]
is coordinate-dependent and, therefore, does not define any
geometric object. For this purpose we are going to relate the
coordinate-dependent quantity $\tQ^{\mu \nu}$ with the external
curvature $Q^a{_b}$ of $S$. We use the form of the metric
introduced in \cite{JKC}:
\begin{equation}\label{gd}
{g}_{\mu\nu} = \left[
\begin{array}{ccccc} n^A n_A & \vline & n_A
& \vline & sM+m^A n_A \\
 & \vline &  & \vline & \\
\hline & \vline &  & \vline &  \\
 n_A & \vline & g_{AB} & \vline & m_A \\
 & \vline &  & \vline & \\
 \hline & \vline &  & \vline &  \\
 sM+m^A n_A & \vline & m_A & \vline &
\left( \frac{M}{N}\right)^2+m^A m_A \\
 \end{array} \right] \ ,
\end{equation}
and
\begin{widetext}
\begin{equation}\label{gu}
{g}^{\mu\nu} = \left[ \begin{array}{ccccc} - \left( \frac 1N
\right)^2 & \vline & \frac {n^A}{N^2} - s\frac {m^A}{M} & \vline &
\frac {s}{M} \\
 & \vline &  & \vline & \\
\hline & \vline &  & \vline &  \\
 \frac {n^A}{N^2} - s\frac {m^A}{M} & \vline &
 {\tilde{\tilde g}}^{AB} - \frac {n^A n^B}{N^2} + s\frac {n^A m^B + m^A
 n^B}{M} & \vline & - s\frac {n^A}{M} \\
 & \vline &  & \vline & \\
 \hline & \vline &  & \vline &  \\
 \frac s{M} & \vline & - s\frac {n^A}{M} & \vline & 0 \\
 \end{array} \right] \ ,
\end{equation}
\end{widetext}
where $M > 0$, $s:=\sgn g^{03}=\pm 1$, $g_{AB}$ is the induced
two-metric on surfaces $\{x^0={\rm const},\; x^3={\rm const} \}$
and ${\tilde{\tilde g}}^{AB}$ is its inverse (contravariant)
metric. Both ${\tilde{\tilde g}}^{AB}$ and $g_{AB}$ are used to
rise and lower indices $A,B = 1,2$ of the two-vectors $n^A$ and
$m^A$.

Formula (\ref{gd}) implies: $\sqrt{|\det g_{\mu\nu}|} = \lambda
M$. Moreover, the object $\Lambda^a$ defined by formula
(\ref{Lambda}), takes the form $\Lambda^a=\lambda X^a$ where
$\lambda$ is given by formula (\ref{lambda}) and $X:=\partial_0 -
n^A\partial_A$. This means that we have chosen the following
degeneracy field: $X^\mu = (1, -n^A , 0)$.

For calculational purposes it is useful to rewrite the
two-dimensional inverse metric ${\tilde{\tilde g}}^{AB}$ in
three-dimensional notation, putting ${\tilde{\tilde g}}^{0a} :=
0$. This object satisfies the obvious identity:
\[
 {\tilde{\tilde g}}^{ac} g_{cb}=\delta^a{_b}-X^a\delta^0{_b} \, .
\]
Hence, the contravariant metric (\ref{gu}) may be rewritten as
follows:
\begin{equation}\label{g2i}
  g^{ab} = {\tilde{\tilde g}}^{ab} -\frac1{N^2} X^aX^b -
 \frac{s}M ( m^a X^b + m^b X^a ) \ ,
\end{equation}
where $m^a:= {\tilde{\tilde g}}^{aB}m_B$, so that $m^0:=0$, and
\[ g^{3\mu} = \frac sM X^\mu \ . \]
It may be easily checked (see, e.g.,~\cite{JKC}, page 406) that
covariant derivatives of the field $X$ {\em along} $S$ are equal
to:
\begin{equation} \label{gradX}
  \nabla_a X = -w_a X - l_{ab}{\tilde{\tilde g}}^{bc}\partial_c
  \  ,
\end{equation}
where
\begin{equation}\label{wu-def}
w_a  := - X^\mu \Gamma^0_{\mu a} \ ,
\end{equation}
and
\begin{equation}\label{el-def}
l_{ab}:=  -g(\partial_b,\nabla_a X) = g(\nabla_a\partial_b, X) =
X_\mu \Gamma^\mu_{ab} \ .
\end{equation}
Since $X$ is orthogonal to $S$, we have $X_a = 0$. Due to
(\ref{gd}), the only non-vanishing component of $X_\mu$ is equal
to $X_3 =sM$. Hence, we have $l_{ab} = sM \Gamma^3_{ab} =sM
A^3_{ab}$ and, consequently,
\begin{equation}\label{A3ab}
  \sqrt{|g|} A^3_{ab} = s\lambda l_{ab} \ .
\end{equation}
Because of identity
\begin{equation}\label{lX}
  X^a l_{ab} = X^a X^c \Gamma_{cab} = \frac 12 X^c X^a g_{ca,b}
  \equiv 0\ ,
\end{equation}
we have also  $l_{ab}X^b=0$ (see \cite{JKC}). Now we are going to
use the metricity condition for the connection $\Gamma$:
\begin{eqnarray}\label{wiaz}
  0  & \equiv & \nabla_a \pi^{3a} =
  \partial_a \pi^{3a} + \pi^{3\mu} \Gamma^a_{\mu a} + \pi^{\mu a}
  \Gamma^3_{\mu a} - \pi^{3a} \Gamma^\mu_{a\mu } \nonumber\\
   & = & \partial_a \pi^{3a} + \pi^{ab} \Gamma^3_{ab} =
   \partial_a \pi^{3a} + \pi^{ab} A^3_{ab} \ .
\end{eqnarray}
Consequently,
\begin{align}\label{dLambda}
  \partial_c \Lambda^c  = \partial_c &\left( s \sqrt{|g|} g^{3c}
  \right) = s{\pi^{3c}}_{,c}=-s\pi^{ab}A^3_{ab}\nonumber\\&=
  -\lambda g^{ab} l_{ab} =
-\lambda {\tilde{\tilde g}}^{ab} l_{ab} = -\lambda  l \ ,
\end{align}
where $l = {\tilde{\tilde g}}^{ab} l_{ab}$.

Now, we want to calculate the component $A^3_{3a} = \Gamma^3_{3a}
- \frac 12 \Gamma^\mu_{\mu a}$. Because
\[
\Gamma^\mu_{\mu a} = \partial_a \ln\sqrt{|g|} =
\partial_a \ln \left( \lambda M \right) \ ,
\]
it is sufficient to calculate $\Gamma^3_{3a}$ according to the
following formula:
\begin{align}
\Gamma^3_{3a}  =&  g^{3c} \Gamma_{c3a} = \frac sM X^c \left(
g_{3c,a}- \Gamma_{3ca} \right) \nonumber \\ =&  \frac sM X^c
g_{3c,a} - X^c g^{0\mu}\Gamma_{\mu ca} +  X^c g^{0b}\Gamma_{bca}
\nonumber \\ =&
 w_a + \frac sM X^c g_{3c,a} + \frac sM  X^b m^c \Gamma_{bca}
 - \frac sM X^c m^b g_{bc,a}  \nonumber \\
 =&  w_a + \frac sM m^c l_{ca}
 \nonumber\\&+ \frac sM \left\{ (X^c g_{3c})_{,a} -
 X^c_{,a} ( g_{3c} - m^b g_{bc} ) \right\} \nonumber \\
 =&  w_a + \frac sM m^c l_{ca} + \frac 1M
  M_{,a}  \ .
\label{A33a}
\end{align}
Finally, we obtain the following identity:
\begin{equation}\label{A3a}
 A^3_{3a}  = w_a +\abar_a +\frac{s}M m^b l_{ba} \ ,
\end{equation}
where $\displaystyle \abar_a := \frac12 \partial_a \ln \left(\frac
M\lambda \right)$.

To express $\tQ$ in terms of $l_{ab}$ and $w_a$, we observe that:
\begin{align}
s {\tQ}^a{_b}& = \lambda \left( g^{ac}l_{cb} -\frac12 \delta^a{_b}
l\right) + \Lambda^a A^3_{3b} - \delta^a{_b} \Lambda^c A^3_{3c} \
,\label{Pl}\\
 s {\tQ}^3{_3}& = -\frac12 \lambda l \ , \label{P33}\\
 s {\tQ}^3{_a}& = 0 \ .\label{P3a}
\end{align}
The missing component $\tQ^a{_3}$ is much more complicated:
\begin{align}
 \tQ^a{_3} & =  \sqrt{|g|} g^{a\beta} A^3_{\beta 3} = \lambda M \left( g^{3a}
A^3_{33}+ g^{ab} A^3_{b3} \right)  = s\Lambda^a A^3_{33}
\nonumber\\+& \lambda M \left\{ {\tilde{\tilde g}}^{ab} +
\frac1{N^2} X^aX^b - \frac{s}M ( m^a X^b + m^b X^a ) \right\}
A^3_{b3}
\end{align}
and depends upon $A^3_{33}$:
\begin{eqnarray}
  A^3_{33} & = & \Gamma^3_{33} - \Gamma^\mu_{3\mu} = - \Gamma^a_{3a}
  =  -\frac12 \left( g^{ab} g_{ab,3} + g^{a3} g_{33,a} \right) \nonumber\\
  &=& -  \partial_3\ln\lambda + \frac{s}M  m^a X^b  g_{ab,3} -
  \frac 12 g^{a3} g_{33,a} \ ,  \label{A333}
\end{eqnarray}
where we have used the identity
\[
 \frac12  {\tilde{\tilde g}}^{ab} g_{ab,3} =
 \partial_3\ln\lambda \ .
\]

We are ready to prove the following
\begin{lemma}
The object ${\tQ}^a{_b}$ is related with $Q^a{_b}$ as follows:
\begin{equation}\label{PQ}
 s {\tQ}^a{_b} = s Q^a{_b} -\frac12 \lambda l \delta^a{_b}+ \Lambda^a
\abar_b -\delta^a{_b} \Lambda^c \abar_c \ ,
\end{equation}
where $\displaystyle \abar_c := \frac12 \partial_c \ln \left(\frac
M\lambda \right)$.
\end{lemma}
\begin{proof}[Proof:]
Using (\ref{Pl}), (\ref{A3a}) and (\ref{g2i}) we obtain:
\begin{align}\label{Pli}
s {\tQ}^a{_b} = &\lambda \left( {\tilde{\tilde g}}^{ac}l_{cb}
-\frac12 \delta^a{_b} l\right) + \Lambda^a w_{b} - \delta^a{_b}
\Lambda^c w_{c}\nonumber\\
 &+ \Lambda^a \abar_b -\delta^a{_b} \Lambda^c
\abar_c \, .
\end{align}
{}From definition (\ref{Q-fund}) and property (\ref{gradX}) one
can check that
\begin{eqnarray}\label{Qli}
 sQ^a{_b} & = & \lambda\delta^a{_b}\nabla_c X^c -\lambda \nabla_b X^a -
 \delta^a{_b} \partial_c \Lambda^c \nonumber\\
 &=& -\lambda\delta^a{_b}(w_cX^c+l) + \lambda (w_bX^a+ {\tilde{\tilde g}}^{ac}
 l_{cb}) + \delta^a{_b} \lambda l \nonumber\\
 &=& \lambda  {\tilde{\tilde g}}^{ac} l_{cb}  +
    \Lambda^a w_b -\delta^a{_b} \Lambda^c w_c
\end{eqnarray}
so we get (\ref{PQ}).
\end{proof}
{\bf Remark:} Formula (\ref{Qli}), together with
$l_{ab}X^b=0=g_{ab}X^b$, gives us the orthogonality condition
$Q^a{_b}X^b =0$ and symmetry of the tensor
$Q_{ab}:=g_{ac}Q^c{_b}$.

Now, we would like to examine the properties of ${\bf G}^{\mu\nu}
= [\tQ^{\mu \nu}]$. {}From continuity of the metric across $S$ we
obtain
\begin{equation}\label{jlab} [l_{ab}] = s M [A^3_{ab}] =
 s M [\Gamma^3_{ab}] = X^c [\Gamma_{cab}]
 = 0 \, .
\end{equation}
On the other hand the jump of $A^3_{3\mu}$ is in general
non-vanishing. {}From ({\ref{A3a}) we have
\begin{equation}\label{jA3a}
 [A^3_{3a}] = [w_a] \, .
\end{equation}
Formulae (\ref{Pl}) -- (\ref{P3a}) and (\ref{jlab}) imply:
\begin{align}\label{jPl}
s [{\tQ}^a{_b}] &=  \Lambda^a [A^3_{3b}] - \delta^a{_b} \Lambda^c
[A^3_{3c}] \nonumber\\& = \Lambda^a [w_{b}] - \delta^a{_b}
\Lambda^c [w_{c}] = s[Q^a{_b}] \ ,
\end{align}
\begin{equation}\label{jP3mu} [{\tQ}^3{_\mu}] = 0 \ .
\end{equation}
Moreover, we have
\begin{align}
 [\tQ^a{_3}] =&
s\Lambda^a \left([A^3_{33}] + s\frac M{N^2} X^b [w_b] -m^b [w_b]
\right)\nonumber\\& + \left(\lambda M {\tilde{\tilde g}}^{ab}
  -s m^a \Lambda^b \right) [w_b] \ .
\end{align}
On the other hand the jump of $A^3_{33}$ may be obtained from
(\ref{A333}):
\begin{equation} \label{jA33}
 [A^3_{33}]= - [\partial_3\ln\lambda] + 2 m^b [w_b] \, ,
 \end{equation}
where we have used
\begin{equation} \label{jwa} [w_a] =  - X^b g^{03}[\Gamma_{3\, ba}]
 =\frac{s}{2M}  X^b [ g_{ab,3}] \, .
\end{equation}
But
\begin{equation}\label{X[w]}
  X^a[w_a] = \frac{s}{2M} [ X^a X^b g_{ab,3}] = 0 \, .
\end{equation}
Hence
\begin{equation} \label{jQa3}
 [\tQ^a{_3}] =   s\Lambda^a \left\{ -[\partial_3\ln\lambda] +m^b [w_b]
\right\} + M \lambda {\tilde{\tilde g}}^{ab}  [w_b] \, .
\end{equation}
Using these results we calculate components of
$[\tQ^{\mu\nu}]={\bf G}^{\mu\nu}$. {}From (\ref{jP3mu}) we can
easily check the property (\ref{G^perp=0})

\begin{eqnarray*}
{\bf G}^{33}=[\tQ^{33}] & = & g^{33}[\tQ^3{_3}] + g^{3b}
[\tQ^3{_b}] = 0 \ , \\ {\bf G}^{3a}=[\tQ^{3a}] & = &
g^{33}[\tQ^a{_3}] + g^{3b} [\tQ^a{_b}]
 = -\frac sM [ X^b Q^a{_b}] = 0 \ ,
\end{eqnarray*}
where we used the property $[\tQ^a{_b}] =  [Q^a{_b}]$ which is
crucial to admit that the object ${\bf G}^a{_b}$ is a well defined
geometric object on $S$. On the contrary, the object ${\bf
G}^{ab}$ is not a geometric object because depends on a choice of
coordinates. This can be seen when we calculate the component
${\bf G}^{00}$:
\begin{align}
{\bf G}^{00} = &[\tQ^{00}] =  g^{03}[\tQ^0{_3}] + g^{0b}
[\tQ^0{_b}] \nonumber\\ =&
\frac{\lambda}M\left(-[\partial_3\ln\lambda] + m^b [w_b]\right)
\nonumber\\ &- s \left(\frac1{N^2} X^b +\frac sM m^b\right)
\lambda [w_b]   =  - \frac{1}M [\partial_3 \lambda] \ .
\end{align}
It may be easily checked (see \cite{JKC}) that the above quantity
transforms in a homogeneous way with respect to coordinate
transformation on $S$. This proves that the components ${\bf
G}^{ab}$ do not define any tensor density on $S$. An independent
argument for this statement may be produced as follows. Begin with
a coordinate system in which we have $X=\partial_0$
(i.e.,~$n^A=0$) and perform the following coordinate
transformation:
\begin{equation}\label{trx0}
 \tilde{x}^{0}= x^0 + b_A x^A \, , \quad \tilde{x}^{A}= x^A
 \, , \quad \tilde{x}^{3}= x^3 \ ,
\end{equation}
where $b_A$ are constant. According to (\ref{gu}) we have:
\begin{align}
\frac s{\tilde M}& = g(d\tilde{x}^{0} , d\tilde{x}^{3})=
g(d{x}^{0} , d{x}^{3}) + b_A g( d x^A , d{x}^{3}) \nonumber\\& =
\frac sM (1 - b_A n^A) =\frac sM \ ,
\end{align}
whence we get ${\tilde M} = M$. Moreover, the new tetrad $({\tilde
X},{\tilde \partial}_{\tilde B} ,
 {\tilde \partial}_{\tilde 3})$ may be calculated as follows:
\begin{align}
{\tilde X} & =   X \ , \label{tr1}
\\
{\tilde \partial}_{\tilde B} & =    \frac{\partial x^0}{\partial
  \tilde{x}^{\tilde{B}}}  \partial_0 +
  \frac{\partial x^A}{\partial
  \tilde{x}^{\tilde{B}}} \partial_A  = \delta_{\tilde{B}}^{\ A}
  \partial_A -  b_{\tilde{B}} X
 \ ,
 \label{tr2}
 \\
 {\tilde \partial}_{\tilde 3} & =  \partial_3 \ .
\end{align}
 This implies ${\tilde \lambda} = \lambda \ ,
$ and, consequently,
\begin{equation}
{\tilde{\bf G}}^{{\tilde 0}{\tilde 0}} = - \frac{1}{\tilde M}
[\partial_{\tilde 3} {\tilde \lambda}] = - \frac{1}M [\partial_3
\lambda] = {\bf G}^{00} \ .
\end{equation}
On the other hand, we have ${\rm d}\tilde{x}^{0}= {\rm d} x^0 +
b_A {\rm d}x^A$ and $\det \left( \frac {\partial x^a}{\partial
\tilde{x}^{\tilde b}}\right)=1$. Hence,
\begin{align*} {\tilde{\bf
G}}^{{\tilde 0}{\tilde 0}} - {\bf G}^{00}&= {\bf G}( {\rm
d}\tilde{x}^{0}, {\rm d}\tilde{x}^{0})-
   {\bf G}( {\rm d}{x}^{0}, {\rm d}{x}^{0})\\
&= 2b_A {\bf G}^{0A} +
   {\bf G}^{AB}b_Ab_B  \, ,
\end{align*}
which does not need to vanish in a generic case.

\section{Gauss-Codazzi equations}

We begin with the Lie derivative of a connection $\Gamma$ with
respect to a vector field $W$ (see \cite{Lie}):
\begin{equation}\label{Lie1}
{\cal L}_W \Gamma^{\lambda}_{\mu\nu} = \nabla_\mu \nabla_\nu
W^\lambda - W^\sigma R^\lambda_{\ \ \nu\mu\sigma} \ .
\end{equation}
For the coordinate field $W=\partial_a$ (i.e., $W^\mu =
\delta^\mu_a$), Lie derivative reduces to the partial derivative:
${\cal L}_W \Gamma^{\lambda}_{\mu\nu} =
\partial_a \, \Gamma^{\lambda}_{\mu\nu} $.
Hence, taking appropriate traces of (\ref{Lie1}) and denoting
$\pi^{\mu\nu}:= \sqrt{|g|} g^{\mu\nu}$ we obtain:
\begin{align*}
{\pi}^{\mu\nu}&  \partial_a \, A^{\alpha}_{\mu\nu} =
(\delta^\alpha_\lambda {\pi}^{\mu\nu} - \delta^\mu_\lambda
{\pi}^{\alpha\nu} )\partial_a \, \Gamma^{\lambda}_{\mu\nu}
\nonumber \\ & =  (\delta^\alpha_\lambda {\pi}^{\mu\nu} -
\delta^\mu_\lambda {\pi}^{\alpha\nu} ) (\nabla_\mu \nabla_\nu
W^\lambda - W^\sigma R^\lambda_{\ \ \nu\mu\sigma}) \nonumber \\ &
=  {\sqrt{|g|}} \left\{ \nabla_\mu (\nabla^\mu W^\alpha -
\nabla^\alpha W^\mu ) + 2 R^{\alpha}_{\ \sigma} W^\sigma \right\}
\nonumber \\
 & =
\partial_\mu \left\{ \sqrt{|g|} (\nabla^\mu W^\alpha - \nabla^\alpha
W^\mu ) \right\} + 2 \sqrt{|g|} R^{\alpha}{_\sigma} W^\sigma
 \ .
\end{align*}
We apply this formula for $\alpha = 3$. This way we have:
\begin{align}
  {\pi}^{\mu\nu}  \partial_a & A^{3}_{\mu\nu}  =
  \partial_\mu \left\{ \sqrt{|g|} (\nabla^\mu W^3 - \nabla^3
W^\mu ) \right\} + 2 {\cal R}^{3}_{\ a} \nonumber
\\
 & =  \partial_b \left\{ \sqrt{|g|} (\nabla^b W^3 - \nabla^3
W^b ) \right\} + 2 {\cal R}^{3}{_a} \ . \label{podst}
\end{align}
where ${\cal R}^{3}{_a}:= \sqrt{|g|} R^{3}{_a} $. But
\[
\nabla_\mu W^\nu = \Gamma^\nu_{a \mu} \ .
\]
Hence:
\begin{align}
  \nabla^b & W^3 - \nabla^3 W^b  \nonumber\\&=
   \frac 12 \left(g^{b\lambda}
  g^{3\mu} - g^{3\lambda}   g^{b\mu} \right) ( g_{\mu
  \lambda,a} + g_{\mu a ,\lambda} - g_{\lambda a , \mu} )
  \nonumber
  \\
  & = g^{b\lambda} g^{3\mu}
  ( g_{\mu a ,\lambda} - g_{\lambda a , \mu} ) = 2 g^{b\lambda}
  \Gamma^3_{\lambda a} - g^{b\lambda} g^{3\mu} g_{\mu \lambda ,
  a}\nonumber
  \\
  & =  2 g^{b\lambda} A^3_{\lambda a} + g^{b3}\Gamma^\mu_{a\mu} + {g^{b3}}_{,a}
  \ , \nonumber
\end{align}
and, consequently,
\begin{equation}\label{nablaW}
   \sqrt{|g|} (\nabla^b W^3 - \nabla^3 W^b ) = 2 \pi^{b\lambda} A^3_{\lambda
   a}+ {\pi^{3b}}_{,a} \ .
\end{equation}
Inserting this to (\ref{podst}) we obtain:
\begin{equation}
\begin{split}\label{fund}
{\cal R}^{3}_{\ a} +  \partial_b \left\{ \pi^{b\lambda}
A^3_{\lambda a}  - \frac 12 \delta^b_a \left( \pi^{\mu\nu}
A^3_{\mu\nu} - {\pi^{3c}}_{,c} \right) \right\}\\ = - \frac 12
{\pi^{\mu\nu}}_{,a} \, A^3_{\mu\nu} \ .
\end{split}
\end{equation}

But
\begin{align}
  -& {\pi^{\mu\nu}}_{,a}  A^3_{\mu\nu}  =  - \left(
  g^{\mu\nu}  \partial_a\sqrt{|g|} +  \sqrt{|g|} \ g^{\mu\alpha} g^{\nu \beta}
  g_{\alpha\beta , a} \right)  A^3_{\mu\nu}\nonumber
  \\
  & =  \left(-\frac 12 g^{\alpha\beta} \pi^{\mu\nu}+
  g^{\alpha\mu} \pi^{\beta\nu}
  \right)    A^3_{\mu\nu}  \, g_{\alpha\beta , a} =
  {\tQ}^{\alpha\beta} g_{\alpha\beta ,  a}\ , \label{pochodna}
\end{align}
where we used definition  (\ref{tQ})), namely
\begin{equation}
\begin{split}
{\tQ}^\mu_{\ \nu} := \sqrt{|g|} \left( g^{\mu \alpha}
A^3_{\alpha\nu} - \frac 12 \delta^\mu_{\ \nu} g^{ \alpha \beta}
A^3_{\alpha\beta} \right)\\ = \pi^{\mu \alpha} A^3_{\alpha\nu} -
\frac 12 \delta^\mu_{\ \nu} \pi^{ \alpha \beta} A^3_{\alpha\beta}
\ .
\end{split}
\end{equation}

 Hence, we obtain the following identity:
\begin{equation}\label{fund1}
{\cal G}^{3}_{\ a} +  \partial_b \left\{ {\tQ}^b{_a}  + \frac 12
\delta^b{_a}  {\pi^{3c}}_{,c}  \right\}  - \frac 12
{\tQ}^{\alpha\beta} g_{\alpha\beta ,  a} \equiv 0 \ .
\end{equation}
To calculate the last term of (\ref{fund1}) we use the following

\begin{lemma}
The following equality holds
\begin{align}
s{\tQ}^{\alpha\beta} g_{\alpha\beta ,a}  =&
  \lambda
 ( g^{be}g^{cd}l_{ed} -\frac12 l g^{bc})
 g_{bc ,a}\nonumber
 \\
& + (\Lambda^b g^{cd} + \Lambda^c g^{bd}
 -\Lambda^d g^{cb})A^3_{3d} g_{bc ,a} \nonumber\\
 &  + 2 s{\tQ}^3{_3} \left( \partial_a \ln M +\frac{s}M m_B
n^B_{,a}\right) \ . \label{Pgfull}
\end{align}
\end{lemma}
\begin{proof}[Proof:]
{}From (\ref{P33}) and (\ref{P3a}) we obtain
\[ \tQ^{33} = 0 \]
and
\[ \tQ^{3b} = g^{3b} \tQ^3{_3} \ , \]
so
\[ {\tQ}^{\alpha\beta} g_{\alpha\beta ,a} = 2 {\tQ}^3{_3} g^{3b}g_{3b,a}
+ {\tQ}^{bc} g_{bc ,a} \, .\] Moreover, from (\ref{gd}) --
(\ref{gu}) we have
\[ g^{3b}g_{3b,a} = \partial_a \ln M +\frac{s}M m_B n^B_{,a} \, .\]
and
\[ {\tQ}^{ab} = \left( \delta^a{_c}g^{bd}+g^{ad}g^{3b}g_{3c} \right)\tQ^c{_d}
+ \frac{X^aX^b}{M^2}\tQ_{33} \, .\] Using (\ref{Pl}) and taking
into account that $X^a X^b g_{ab,c}=0$ we get
\begin{align} \label{Pgabc}
 s{\tQ}^{bc} g_{bc ,a} =&  \lambda
 ( g^{be}g^{cd}l_{ed} -\frac12 l g^{bc})
 g_{bc ,a}\nonumber\\
& + \Lambda^b A^3_{3d}g^{cd} + \Lambda^c A^3_{3d}g^{bd}
 -\Lambda^d A^3_{3d}g^{cb}  g_{bc ,a}
\end{align}
and finally
\begin{align}
s{\tQ}^{\alpha\beta} g_{\alpha\beta ,a}  = &  \lambda
 ( g^{be}g^{cd}l_{ed} -\frac12 l g^{bc})
 g_{bc ,a}\nonumber\\
& + (\Lambda^b g^{cd} + \Lambda^c g^{bd}
 -\Lambda^d g^{cb})A^3_{3d}  g_{bc ,a}\nonumber \\
 &  + 2 s{\tQ}^3{_3} \left( \partial_a \ln M +\frac{s}M m_B
n^B_{,a}\right) \label{Pgfull1} \, .
\end{align}
\end{proof}

Now, the proof of (\ref{G-C}) is roughly a straightforward
calculation starting from equation (\ref{fund1}) and consequent
reexpressing all ingredients in terms of the connection objects
$l_{ab}$, $w_a$ and the metric objects $M, m^A, N, X^a, g_{ab}$
describing the 4-dimensional metric $g_{\mu\nu}$. It turns out
that the terms containing $M, N$ and $m^A$ drop out. Inserting
(\ref{PQ}) and (\ref{Pgfull}) into (\ref{fund1}) and using
(\ref{A3a}), (\ref{P33}), and (\ref{g2i}), we obtain:
\begin{align}
s{\cal G}^{3}_{\ a}  = & -s\partial_b \left\{ {\tQ}^b{_a}  + \frac
12 \delta^b{_a}  {\pi^{3c}}_{,c}  \right\}  + s \frac 12
{\tQ}^{\alpha\beta} g_{\alpha\beta , a} \nonumber \\ \nonumber =&
\partial_b\left\{ -s{Q}^{b}{_a} + \delta^b{_a} \lambda l - \Lambda^a
\abar_b +\delta^a{_b} \Lambda^c \abar_c\right\}\\& -\frac12
\lambda l (\partial_a \ln M +\frac{s}M m_B n^B_{,a})\nonumber\\
\nonumber &  +\frac12 g_{bc , a} \left( \Lambda^b g^{cd} +
\Lambda^c g^{bd}
 -\Lambda^d g^{cb} \right)\\
&\qquad\times
 \left( w_d +\abar_d +\frac{s}M m^B l_{Bd}\right )\nonumber
 \\ \nonumber
&  +\frac12 \lambda g_{bc , a} \left( g^{be}g^{cd}l_{ed} -\frac12
l g^{bc}
 \right) \\
 =& -s\partial_b {Q}^{b}{_a} +\frac 12 s{Q}^{bc}
g_{bc , a} +\lambda \partial_a l \ ,\label{fund2}
\end{align}
where we have used formula
\[
    sQ^{ab}= \lambda  {\tilde{\tilde g}}^{ac}
    {\tilde{\tilde g}}^{bd} l_{cd} +
    ( \Lambda^a {\tilde{\tilde g}}^{bc} +
    \Lambda^b {\tilde{\tilde g}}^{ac}
    - {\tilde{\tilde g}}^{ab} \Lambda^c) w_c \, .
\]
Formula (\ref{fund2}) is equivalent to (\ref{G-C}) if we use
(\ref{dLambda}), and keep in mind the ``gauge'' condition
$X(x^0)=1$, used thoroughly in this proof.

\section{Proof of lemma  (\ref{lem1}) and examples of
invariant Lagrangians} \label{proof-matter-Lagrangian}

Since the matter Lagrangian (\ref{L1}) is an invariant scalar
density, its value may be calculated in any coordinate system. For
purposes of the proof let us restrict ourselves to local
coordinate systems $(x^a)$ on $S$ which are compatible with the
degeneracy of the metric, i.e.,~such that $X:=\partial_0$ is
null-like.

Suppose that $(x^a)$ and $(y^a)$ are two such local systems in a
neighborhood of a point $x \in S$. Suppose, moreover, that both
vectors $\partial_0$ coincide. It is easy to see that these
conditions imply the following form of the transformation between
the two systems:
\begin{eqnarray}\label{trans:xA-yA}
  y^A &=& y^A ( x^B ) \ , \\
  y^0 &=& x^0 + \psi (x^A) \ . \label{trans:x0-y0}
\end{eqnarray}
Three-dimensional Jacobian of such a transformation is equal to
the two-dimensional one: $\det ( \partial y^A / \partial x^B ) $.
Observe that the two-dimensional part $g_{AB}$ of the metric
$g_{ab}$ transforms according to the same two-dimensional matrix
and, whence, its determinant $\lambda$ gets multiplied by the same
two-dimensional Jacobian when transformed from $(x^a)$ to $(y^a)$.
So does also the volume $v_X$. This means that the function
\begin{equation}\label{tilde-f}
  f := \frac L{v_X} \ ,
\end{equation}
does not change value during such a transformation. A priori, we
could have:
\begin{equation}\label{tiled-f-dep}
  f = f (z^K;{z^K}_0;{z^K}_A;  g_{ab}) \ ,
\end{equation}
but we are going to prove that, in fact, it {\em cannot} depend
upon derivatives ${z^K}_A$. For this purpose consider new
coordinates:
\begin{eqnarray}\label{trans:x-y-1}
  y^A &=& x^A \ , \\
  y^0 &=& x^0 - \epsilon_1 x^1  - \epsilon_2 x^2 \ .
\end{eqnarray}
This implies that
\[
\frac {\partial}{\partial y^A} = \frac {\partial}{\partial x^A} +
\epsilon_A \frac {\partial}{\partial x^0} \ .
\]
Passing from $(x^a)$ to $(y^a)$, the value of ${z^K}_A$ will be
thus replaced by ${z^K}_A + \epsilon_A {z^K}_0$, whereas the
remaining variables of the function (\ref{tiled-f-dep}) (and also
its value) will remain unchanged. This implies the following
identity:
\begin{equation}\label{tilde-f-niez}
  f(z^K;{z^K}_0;{z^K}_A;  g_{ab}) =
  f(z^K;{z^K}_0;{z^K}_A + \epsilon_A
  {z^K}_0;  g_{ab}) \ ,
\end{equation}
which must be valid for any configuration of the field $z^K$. Such
a function cannot depend upon ${z^K}_A $! But in our coordinate
system we have ${z^K}_0 ={z^K}_a  X^a = {\cal L}_X z^K $. Thus, we
have proved that
\begin{equation}\label{tiled-f-dep-n}
  f = f(z^K;{\cal L}_X z^K  ;  g_{ab}) \ .
\end{equation}
Relaxing condition (\ref{trans:x0-y0}) and admitting arbitrary
time coordinates $y^0$, we easily see that the dependence of
(\ref{tiled-f-dep-n}) upon its second variable must annihilate the
(homogeneous of degree minus one) dependence of the density $v_X$
upon the field $X$ in formula (\ref{lagr-form}). This proves that
$f$ must be homogeneous of degree one in ${\cal L}_X z^K  $.

As an example of an invariant Lagrangian consider a theory of a
light-like ``elastic media'' described by material variables
$z^A$, $A=1,2$, considered as coordinates in a two-dimensional
material space $Z$, equipped with a Riemannian ``material metric''
$\gamma_{AB}$. Moreover, take a scalar field $\xi$. Then for
numbers $\alpha $ and $\beta > 0$, satisfying identity
$2\alpha+\beta=1$, and for any function $\psi$ of one variable,
the following Lagrangian density:
\begin{equation}
L=\lambda \  \psi (\xi) \left(X^a {\partial z^K \over
\partial x^a} X^b {\partial z^L \over \partial x^b} \,
\gamma_{KL}(z^A)\right)^\alpha \left(X^c {\partial \xi \over
\partial x^c} \right)^{\beta}\ ,
\end{equation}
fulfills properties listed in Lemma (\ref{lem1}) and, therefore,
is invariant. If  $\psi$ is constant, a possible physical
interpretation of the variable $\xi$ as a ``thermodynamical
potential'', may be found in \cite{KSG}.

\section{Reduction of the generating formula}

\subsection{Proof of formulae (\ref{redukcja-0}) and (\ref{redukcja-3})}

\label{ap-redukcja}

We reduce the generating formula with respect to constraints
implied by identities $\nabla_{k}{\pi}^{0k}=0$ and
$\nabla_{k}{\pi}^{00}=0$. In fact, expressing the left-hand sides
in terms of ${\pi}^{\mu\nu}$ and $A_{\mu\nu }^{0}$ we immediately
get the following constraints:
\begin{align}
A_{00}^{0}  &  =\frac{1}{{\pi}^{00}}\left(  \partial_{k}{\pi}^{0k}+A_{kl}%
^{0}{\pi}^{kl}\right)  \ ,\label{A000}\\
A_{0k}^{0}  &  =-\frac{1}{2{\pi}^{00}}\left(  \partial_{k}{\pi}^{00}%
+2A_{kl}^{0}{\pi}^{0l}\right)  \ . \label{A00k}%
\end{align}
It is easy to see that they imply the following formula:
\begin{align}
{\pi}^{\mu\nu}\delta A_{\mu\nu}^{0}  &  ={\pi}^{kl}\delta
A_{kl}^{0}+2{\pi }^{0k}\delta A_{0k}^{0}+{\pi}^{00}\delta
A_{00}^{0}\nonumber\\
&  =-\frac{1}{16\pi}g_{kl}\delta P^{kl}+\partial_{k}\left(  {\pi}^{00}%
\delta\left(  \frac{{\pi}^{0k}}{{\pi}^{00}}\right)  \right)  \ ,
\label{pi-dA0}%
\end{align}
where we have denoted
\begin{align}
P^{kl}:=  &  \sqrt{\det g_{mn}}\
(K{\tilde{g}}^{kl}-K^{kl})\label{Pkl}\\
K_{kl}:=  &  -\frac{1}{\sqrt{|g^{00}|}}{\Gamma}_{kl}^{0}=-\frac{1}%
{\sqrt{|g^{00}|}}A_{kl}^{0}\nonumber
\end{align}
and ${\tilde{g}}^{kl}$ is the 3-dimensional inverse with respect
to the induced metric $g_{kl}$ on $V$.

Let us exchange now the role of $x^{3}$ and $x^{0}$ . Identities
(\ref{A000}) and (\ref{A00k})
 become
constraints on the boundary of the world-tube $\partial{\cal U}$:
\begin{align}
A_{33}^{3}  &  =\frac{1}{{\pi}^{33}}\left(  \partial_{a}{\pi}^{3a}+A_{ab}%
^{3}{\pi}^{ab}\right)  \ ,\\
A_{3a}^{3}  &  =-\frac{1}{2{\pi}^{33}}\left(  \partial_{a}{\pi}^{33}%
+2A_{ab}^{3}{\pi}^{3b}\right)  \ .
\end{align}
They imply:
\begin{align}
{\pi}^{\mu\nu}\delta A_{\mu\nu}^{3}  &  ={\pi}^{ab}\delta
A_{ab}^{3}+2{\pi }^{3a}\delta A_{3a}^{3}+{\pi}^{33}\delta
A_{33}^{3}\nonumber\\
&  =-\frac{1}{16\pi}g_{ab}\delta Q^{kl}+\partial_{a}\left(  {\pi}^{33}%
\delta\left(  \frac{{\pi}^{3a}}{{\pi}^{33}}\right)  \right)  \ ,
\end{align}
where we have denoted
\begin{equation}
\mathcal{Q}^{ab}=\sqrt{|\det{g_{cd}}|}\left(
L\tilde{g}^{ab}-L^{ab}\right) \ ,\
L_{ab}=-\frac{1}{\sqrt{g^{33}}}\Gamma_{ab}^{3}\ , \label{qkl}
\end{equation}
and $\ \tilde{g}^{ab}$ is the 3-dimensional inverse with respect
to the induced metric $g_{ab}$ on the world-tube.

\subsection{ Proof of formula \ref{lambda-alpha}}
\label{ap-lambda-alpha}
Write the right hand side as follows:%
\begin{equation}
\pi^{00}\delta\left(  \frac{\pi^{03}}{\pi^{00}}\right)
+\pi^{33}\delta\left( \frac{\pi^{30}}{\pi^{33}}\right)
=2\sqrt{|\pi^{00}\pi^{33}|}\delta\frac
{\pi^{30}}{\sqrt{|\pi^{00}\pi^{33}|}}\ ,
\end{equation}
and
\begin{equation}
2\sqrt{|\pi^{00}\pi^{33}|}=\frac{2}{16\pi}\sqrt{|g|}\sqrt{|g^{00}g^{33}%
|}=\frac{1}{8\pi}\frac{\sqrt{\det g_{AB}}}{\sqrt{1+q^{2}}}\ .
\end{equation}
This automatically implies
\begin{equation}
\pi^{00}\delta\left(  \frac{\pi^{03}}{\pi^{00}}\right)
+\pi^{33}\delta\left(
\frac{\pi^{30}}{\pi^{33}}\right)  =\frac{\lambda}{8\pi}\frac{\delta q}%
{\sqrt{1+q^{2}}}\ =\frac{\lambda}{8\pi}\delta\alpha.
\end{equation}

\subsection{ Proof of formula \ref{pe-lambda1}}
\label{ap-pe-lambda}

To prove (\ref{pe-lambda1}), consider first the following
identity:
\begin{equation}\label{pe-dot-de}
\int_Vg_{kl}\dot{P}^{kl}=-\int_V D+\int_{ V}\partial_k
\left(\pi^{00}\partial_0 \left(\frac{\pi^{0k}}{\pi^{00}}\right)
\right)\ ,
\end{equation}
where we denote:
\begin{align*}
D:&=-\frac{1}{16\pi}g_{kl}{\dot{P}}^{kl}+\partial_{k}\left(
{\pi}^{00} \partial_0 \left( \frac{{\pi}^{0k}}{{\pi}^{00}}\right)
\right) \\&={\pi}_{\lambda}^{\ \mu
\nu0}\partial_{0}\Gamma_{\mu\nu}^{\lambda}={\pi}_{\lambda}^{\
\mu\nu 0}\mathcal{L}_{X}\Gamma_{\mu\nu}^{\lambda}\ ,
\end{align*}
with $X=\frac{\partial}{\partial x^{0}}$, i.e.,
$X^{\mu}=\delta_{0}^{\mu}$ and $\mathcal{L}_{X}$ being the Lie
derivative with respect to the field $X$:
\[
\mathcal{L}_{X}\Gamma_{\mu\nu}^{\lambda}=\nabla_{\mu}\nabla_{\nu}X^{\lambda
}-X^{\sigma}R_{\ \ \nu\mu\sigma}^{\lambda}%
\]
(due to Bianchi identities the right hand side is automatically
symmetric with respect to lower indices). Hence
\begin{align}
D:&=
(\delta_{\lambda}^{0}{\pi}^{\mu\nu}-\delta_{\lambda}^{\mu}{\pi}^{0\nu
})(\nabla_{\mu}\nabla_{\nu}X^{\lambda}-X^{\sigma}R_{\ \ \nu\mu\sigma}%
^{\lambda})\nonumber\\ &  =\frac{\sqrt{|g|}}{16\pi}\left\{
\nabla_{\mu}(\nabla^{\mu}X^{0}-\nabla ^{0}X^{\mu})+2R_{\
\sigma}^{0}X^{\sigma}\right\}  =\nonumber\\
&  =\frac{1}{16\pi}\left\{  \partial_{k}\left(  \sqrt{|g|}(\nabla^{k}%
X^{0}-\nabla^{0}X^{k})\right)  +2\sqrt{|g|}R_{\ 0}^{0}\right\}  \
.
\end{align}
The covariant derivative $\nabla_{\mu}$ has been replaced in the
last equation by the partial derivative $\partial_{\mu}$, because
they both coincide when acting on antisymmetric, covariant
bivector densities. We use also identity
\begin{equation}
\nabla^{\mu}X^{\nu}=g^{\mu\lambda}X^{\sigma}\Gamma_{\sigma\lambda}^{\nu
}=g^{\mu\lambda}\Gamma_{0\lambda}^{\nu}\ .
\end{equation}
which finally implies:
\begin{equation}
\begin{split}
\int_{V}D   = \frac{1}{16\pi}\int_V\partial_\nu\left(\sqrt{| g|}
\left(g^{\nu\mu}\Gamma^0_{0\mu}-g^{0\mu}\Gamma^\nu_{0\mu}\right)\right)\\
+\frac{1}{8\pi}\int _{V}\sqrt{|g|}R_{\ 0}^{0} \ .
\end{split}
\end{equation}
$D$ is regular, because singular expressions contained in its
definition cancel out, as implied by equation (\ref{pe-dot-de}).
Hence, we treat $D$ as a regular expression, and there is no need
to integrate it in distributional sense. Hence we have:
\begin{align}
&\int_Vg_{kl}\dot{P}^{kl}=-\frac{1}{8\pi}\int_V\sqrt{| g|} R^0{_0}
\nonumber
\\
&- \frac{1}{16\pi}\int_{\partial V}\sqrt{| g|} \left(
g^{3\mu}\Gamma^0_{0\mu} g^{0\mu}\Gamma^3_{0\mu}-
\pi^{00}\partial_0\left(\frac{\pi^{03}}{\pi^{00}}\right) \right) \
.
\end{align}
{}From definition of $\alpha$ we also have:
\begin{equation}
\lambda\dot{\alpha}=8\pi\left(
\pi^{00}\partial_0\left(\frac{\pi^{03}}{\pi^{00}}\right)+
\pi^{33}\partial_0\left(\frac{\pi^{30}}{\pi^{33}}\right) \right)
\end{equation}
Using the above formula we may write
\begin{align}\label{pe-lambda}
-&\frac{1}{16\pi}\int_{V}\left(  g_{kl}\dot P^{kl}\right) +\frac
{1}{8\pi}\int_{\partial V}\lambda\dot{\alpha}  =
\frac{1}{8\pi}\int_V\sqrt{| g|} R^0{_0} \nonumber \\
&+\frac{1}{16\pi}\int_{\partial V}\sqrt{| g|}
\left(g^{3\mu}\Gamma^0_{0\mu}-g^{0\mu}\Gamma^3_{0\mu}+
g^{33}\partial_0\left(\frac{\pi^{30}}{\pi^{33}}\right)
 \right) \ .
\end{align}
The left-hand side of the above equation is regular, but on the
right-hand side singular terms like $\frac{1}{8\pi}\int_V\sqrt{|
g|} R^0{_0}$ and $\frac{1}{16\pi}\int_{\partial V}\sqrt{| g|}
\left(g^{3\mu}\Gamma^0_{0\mu}-g^{0\mu}\Gamma^3_{0\mu}\right)$
arise. The latter quantity, although it is a boundary term,
origins from the volume term
$\frac{1}{16\pi}\int_V\partial_\nu\left(\sqrt{| g|}
\left(g^{\nu\mu}\Gamma^0_{0\mu}-
g^{0\mu}\Gamma^\nu_{0\mu}\right)\right)$ {\em via} Stokes theorem.
{}From derivatives in $x^3$-direction there come singular terms,
which cancel out the singular part of $R^0{_0}$, giving regular
expression as a final result.

We may rewrite expressions in (\ref{pe-lambda})  in terms of the
quantity $\mathcal{Q}^{ab}$ (defined by (\ref{qkl}))
\begin{eqnarray}
& & \frac{1}{16\pi}\int_{\partial V}\sqrt{|g|}\left(  g^{3\mu}%
\Gamma_{\mu0}^{0}-g^{0\mu}\Gamma_{\mu0}^{3}+
g^{33}\partial_0\left(\frac{\pi^{30}}{\pi^{33}}\right) \right)
  =  \nonumber \\ & &
 \frac{1}{16\pi}\int_{\partial V}\left(  \mathcal{Q}^{AB}g_{AB}%
-\mathcal{Q}^{00}g_{00}\right)
\end{eqnarray}
what completes the proof of formula (\ref{pe-lambda1}).

\section{Proof of the constraint (\ref{plk})}\label{dPk}

Using the decomposition (\ref{gd}), (\ref{gu}) of the metric, one
can express vector $n$ orthonormal to $V_t$ as follows:
\[
n=\frac{1}{N}\left(\partial_0
-n^A\partial_A+s\frac{N^2}{M}m^A\partial_A-s\frac{N^2}{M}\partial_3\right)\
.
\]
Choosing $X=\partial_0 -n^A\partial_A$, we have:
\begin{equation}
\frac1N X=s\frac{N}{M}(\partial_3-m^A\partial_A)+ n \ .
\end{equation}
Consequently, we can rewrite the left-hand side of (\ref{GXt}) as
follows:
\begin{equation}
\frac1N {\cal G}^0{_\mu}X^\mu =s\frac{N}{M}{\cal
G}^0{_3}-s\frac{N}{M}m^A {\cal G}^0{_A}+ {\cal G}^0{_\mu} n^\mu \,
.
\end{equation}
Expressing ${\cal G}^0{_\mu}$ in terms of the canonical
ADM~momentum $P_{kl}$ (equations (\ref{GCww}) and (\ref{GCws})),
equation (\ref{GXt}) takes the form:
\begin{align} \nonumber
0=\frac1N{\cal G}^0{_\mu}X^\mu =& s\frac{N}{M}(P_3{^k}_{|k}-m^A
P_A{^k}_{|k}) \nonumber\\& + \frac12\left(
\sgth\stackrel{(3)}{R}-\left(P^{kl}P_{kl}-\frac{1}{2}
P^2\right)\sgt \right) \ .
\end{align}
Equations (\ref{wwd}) and (\ref{wsd}) give us the following
result:
\begin{equation} \label{ufws}
s\frac NM\left([P_3{^3}]-m^A [P_A{^3}]\right) +[\lambda k] = 0 \ .
\end{equation}
Due to (\ref{gd}), one can express the three-dimensional inverse
metric $\tilde{g}^{kl}$ as follows:
\begin{equation}
\tilde{g}^{kl}=\left(\frac{N}{M}\right)^2\left[
\begin{array}{ccc} \biggl(\left(\frac{M}{N}\right)^2+m^A
m_A\biggr)\tilde{\tilde{g}}^{AB} & \vline &-m^A\\ & \vline & \\
\hline & \vline & \\ -m^A & \vline & 1
\end{array}
\right] \ .
\end{equation}
The above form of $\tilde{g}^{kl}$ can be used to rewrite the
canonical momentum part of (\ref{ufws}):
\begin{align}
s\frac NM&\left([P_3{^3}]-m^A [P_A{^3}]\right)\nonumber\\ &=s\frac
MN [P^{33}]=\frac{\sgth}{\lambda}[P^{33}]=
\left[\frac{P^{33}}{\sqrt{\gt^{33}}} \right] \ ,
\end{align}
and finally we obtain the constraint (\ref{plk}).


\end{document}